\documentclass[sigconf]{acmart}
\usepackage[english]{babel}
\usepackage{blindtext}
\usepackage{mathrsfs}
\usepackage{subfigure}
\usepackage{multirow}
\usepackage{graphicx} 
\usepackage{subcaption}
\usepackage[T1]{fontenc}
\usepackage{aecompl}
\usepackage{booktabs}
\usepackage{enumitem}
\usepackage{tikz}
\usepackage[ruled,vlined,linesnumbered]{algorithm2e}

\AtBeginDocument{%
  }

\acmYear{2024}\copyrightyear{2024}
\setcopyright{rightsretained}
\acmConference[ACM SIGCOMM '24]{ACM SIGCOMM 2024 Conference}{August 4--8, 2024}{Sydney, NSW, Australia}
\acmBooktitle{ACM SIGCOMM 2024 Conference (ACM SIGCOMM '24), August 4--8, 2024, Sydney, NSW, Australia}
\acmDOI{10.1145/3651890.3672258}
\acmISBN{979-8-4007-0614-1/24/08}

\begin{document}
\begin{sloppypar}
\title{FIGRET: Fine-Grained Robustness-Enhanced Traffic Engineering}


\author{Ximeng Liu}
\affiliation{%
  \institution{Shanghai Jiao Tong University}
  \city{}
  \country{}
}

\author{Shizhen Zhao}\authornote{Shizhen Zhao is the corresponding author.}
\affiliation{%
  \institution{Shanghai Jiao Tong University}
  \city{}
  \country{}}

\author{Yong Cui}
\affiliation{%
  \institution{Tsinghua University}
  \city{}
  \country{}
}

\author{Xinbing Wang}
\affiliation{%
 \institution{Shanghai Jiao Tong University}
 \city{}
 \state{}
 \country{}}

\renewcommand{\shortauthors}{Liu et al.}

\begin{abstract}
Traffic Engineering (TE) is critical for improving network performance and reliability. A key challenge in TE is the management of sudden traffic bursts. Existing TE schemes either do not handle traffic bursts or uniformly guard against traffic bursts,
thereby facing difficulties in achieving a balance between normal-case performance and burst-case performance. To address this issue, we introduce FIGRET, a Fine-Grained Robustness-Enhanced TE scheme. FIGRET offers a novel approach to TE by providing varying levels of robustness enhancements, customized according to the distinct traffic characteristics of various source-destination pairs. By leveraging a burst-aware loss function and deep learning techniques, FIGRET is capable of generating high-quality TE solutions efficiently. Our evaluations of real-world production networks, including Wide Area Networks and data centers, demonstrate that FIGRET significantly outperforms existing TE schemes. Compared to the TE scheme currently deployed in Google’s Jupiter data center networks, FIGRET achieves a 9\%-34\% reduction in average Maximum Link Utilization and improves solution speed by $35\times$-$1800 \times$. Against DOTE, a state-of-the-art deep learning-based TE method, FIGRET substantially lowers the occurrence of significant congestion events triggered by traffic bursts by 41\%-53.9\%  in topologies with high traffic dynamics.
\end{abstract}


\maketitle

\section{introduction}
With the exponential growth in network traffic, both data center networks \cite{poutievski2022jupiter,alizadeh2014conga,benson2011microte,al2010hedera,teh2020couder,cao2021trod,chen2018auto} and Wide Area Networks (WANs) \cite{jiang2009cooperative,kandula2005walking,kandula2014calendaring,kumar2018semi,elwalid2001mate,fortz2000internet,zhong2021arrow,wang2006cope,roughan2003traffic,applegate2003making,suchara2011network,xu2023teal} are increasingly dependent on Traffic Engineering (TE) to optimize network performance. TE, typically enabled by a centralized controller in Software-Defined Networking (SDN) \cite{jain2013b4,poutievski2022jupiter,akyildiz2014roadmap, agarwal2013traffic,zaicu2021helix,bogle2019teavar,hong2013achieving,kumar2015bwe,liu2014traffic}, periodically solves optimization problems to efficiently route traffic across network paths, and then translates these solutions into router configurations.

A major challenge in TE is managing sudden traffic bursts. Considering the latency introduced by the central controller in collecting traffic demands, computing new TE solutions, and updating forwarding rules, A TE system should pre-compute network configurations based on historical data prior to the arrival of actual traffic flows. Yet, the inherently dynamic and unpredictable nature of real network traffic poses significant forecasting difficulties \cite{teixeira2005bgp,xu2005profiling}. Inadequate preparation for traffic bursts may result in severe network congestion, leading to prolonged delays, high packet loss rates, and diminished network throughput \cite{wang2006cope}. Thus, enhancing the robustness against unexpected traffic bursts is essential.

Existing burst-aware TE schemes usually handle traffic bursts at the cost of compromised normal-case network performance. Oblivious routing ~\cite{applegate2003making} optimizes TE solutions for the worst-case scenario across all traffic demands. While this approach offers the highest level of robustness against traffic bursts, it tends to result in highly suboptimal performance for non-burst traffic patterns, which constitute the majority of traffic over time. As an improvement, Cope \cite{wang2006cope} focuses on optimizing the predicted traffic demands while providing a worst-case performance guarantee. However, offering such a worst-case guarantee may still be an overkill because some traffic patterns may never occur. In addition, the computational complexity of COPE is also much higher. Consequently, a new class of TE methods \cite{wang2006cope,poutievski2022jupiter,teh2020couder,teh2022enabling} are proposed. These TE methods do not offer guarantee in the entire traffic pattern space, but instead enhance robustness by directly limiting the routing weights of different paths. Specifically, COUDER \cite{teh2022enabling} introduces the \emph{path sensitivity} metric to evaluate the impact of burst traffic on each path, and enhances robustness by minimizing the maximum sensitivity across all paths. Similarly, Google’s optical data centers adopt a hedging mechanism in TE\cite{poutievski2022jupiter}, which enhances robustness by confining the path sensitivity under a predetermined upper limit. However, these methods may still impact TE performance under non-burst traffic scenarios because they force traffic to spread across multiple (potentially longer) paths rather than taking the best path, even if the traffic between certain source-destination pairs is stable.

The limitations observed in the previous methods can be attributed to their uniform treatment against traffic bursts.
In practice, network traffic exhibits different characteristics among different source-destination pairs. While some pairs may frequently encounter traffic bursts, others could remain remarkably stable. For consistently stable traffic, prioritizing robustness becomes unnecessary and may even hurt performance.


Building upon this observation, we design FIGRET (Fine-Grained Robustness-Enhanced Traffic Engineering). FIGRET's \textit{key insight} lies in a fine-grained customized robustness enhancement strategy, tailored for every source-destination pair based on their traffic characteristics. It applies relaxed robustness requirements to source-destination pairs with stable traffic and enforces stricter requirements for those prone to bursts. Similar to \cite{teh2020couder}, FIGRET also uses the path sensitivity metric to enhance robustness against traffic bursts. On top of this, FIGRET customizes the path sensitivity constraints according to the network topology and the traffic characteristics of different source-destination pairs. This strategy allows FIGRET to achieve fine-grained robustness enhancement and ensure a satisfactory balance in TE performance between both normal and burst traffic scenarios.

Having proposed the FIGRET formulation, the next challenge is to efficiently compute a TE solution. At first glance, FIGRET's formulation can be directly solved using linear programming. However, this method has two shortcomings. First, directly solving FIGRET requires a predicted traffic matrix. However, due to the existence of highly bursty source-destination pairs, finding an appropriate traffic prediction is difficult. Second, linear programming involves high computational complexity and may not scale to large networks.
To address these issues, FIGRET leverages a deep neural network to accelerate TE computation. Similar to DOTE~\cite{perry2023dote}, FIGRET directly maps a history of traffic patterns to a routing weight configuration, thus eliminating the need of traffic prediction. To handle the path sensitivity constraints, FIGRET adds an additional term in its loss function to capture the customized robustness requirement.


We conduct a comprehensive evaluation of FIGRET. This evaluation utilizes publicly available WAN datasets \cite{uhlig2006providing}, as well as Data Center PoD-level and ToR-level topologies and traffic data \cite{alizadeh2013pfabric,roy2015inside}. The data encompasses topologies ranging from dozens to hundreds of nodes, with the corresponding traffic data exhibiting various characteristics, including profiles with low, moderate, and high burstiness.
Through our evaluation, we find that FIGRET consistently delivers high-quality TE solutions across a variety of topologies. Compared to the TE system in Google's production data center \cite{poutievski2022jupiter}, FIGRET achieves an average Max Link Utilization (MLU) reduction of 9\%-34\% across different topologies and improves solution speed by $35\times$-$1800 \times$. In comparison with the state-of-the-art Deep Learning-based TE system DOTE \cite{perry2023dote}, specifically designed for MLU optimization, FIGRET achieves notable improvements in two topologies with bursty traffic data. It reduced the average Max Link MLU by 4.5\% and 5.3\% while simultaneously decreasing the number of significant congestion incidents caused by traffic bursts by 41\% and 53.9\%. Meanwhile, in topologies with stable traffic data, FIGRET performs at least as well as DOTE, despite of the additional consideration of robustness. Finally, we numerically interpret FIGRET's superior performance. Our code is available at \cite{Figret2024}.

This work does not raise any ethical concerns.
\section{motivation and key insights}
\label{section: motivation}
\subsection{Necessity of managing bursts}
\label{section: necessity of managing bursts}
Traffic engineering has been adopted in both Wide Area Networks (e.g., Google’s B4 \cite{jain2013b4} and Microsoft’s SWAN \cite{hong2013achieving}) and data centers (e.g., Google’s optical data center network \cite{poutievski2022jupiter}) to improve network utilization and prevent network congestion.

\begin{figure*}[!ht]
    \centering
    \begin{minipage}{0.62\textwidth}
        \subfigure[WAN GEANT, 500 demands collected every 15 minutes]{
        \label{fig:GEANT burst}
        \includegraphics[scale=0.2]{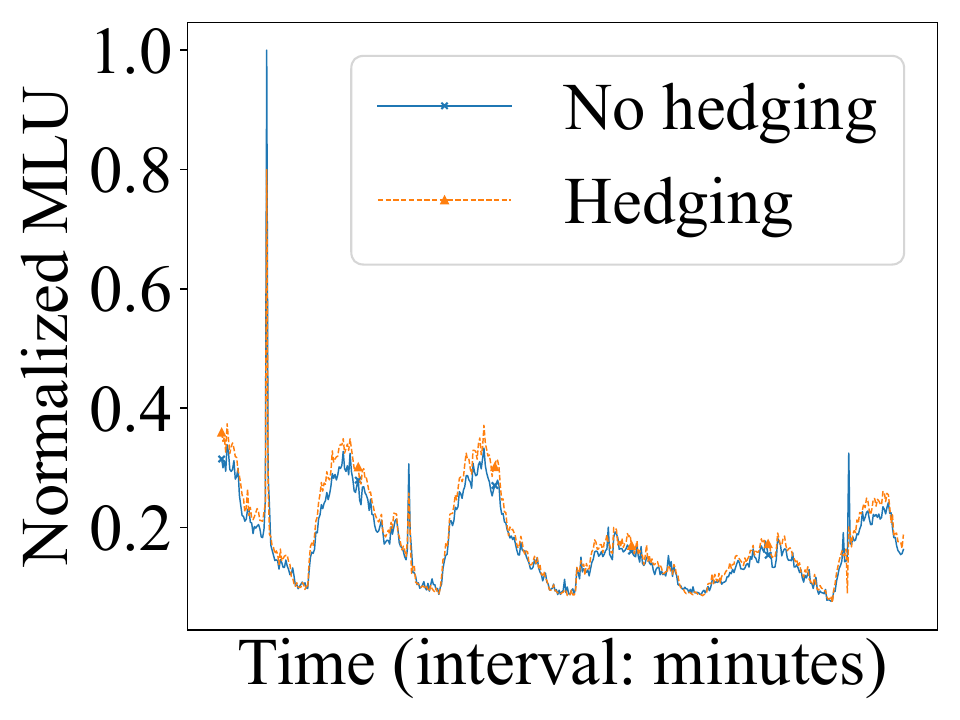}
        }
        \subfigure[PoD-level, Meta data center, 500 demands collected every 1 minute]{
        \label{fig:Facebook_pod_a burst}
        \includegraphics[scale=0.2]{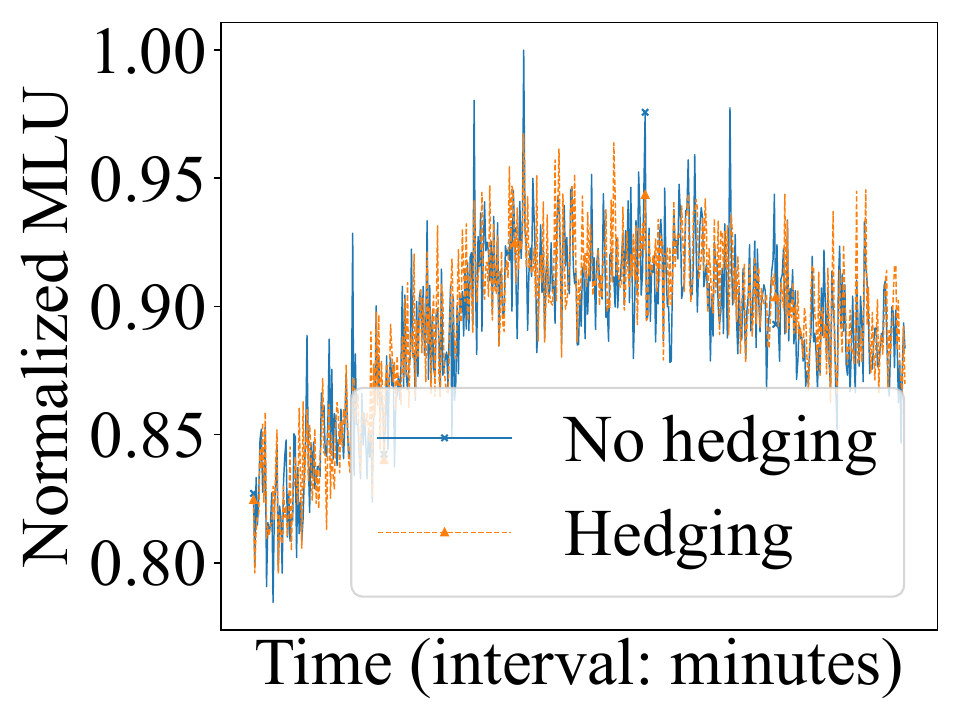}
        }
        \subfigure[ToR-level, Meta data center, 500 demands collected every 1 minute]{
        \label{fig:Facebook_tor_a burst}
        \includegraphics[scale=0.2]{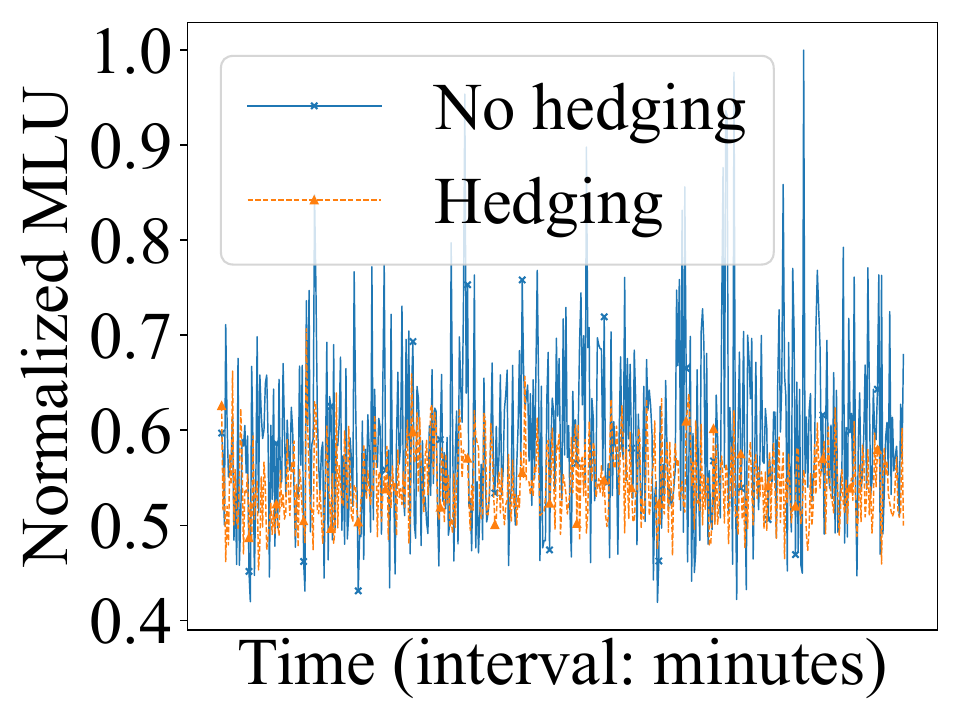}
        }
        \caption{Comparing the impact of using anti-burst Hedging mechanism on maximum link utilization (MLU), with MLU values normalized to the maximum MLU.}
        \label{fig: anti-burst motivation}

        \subfigure[WAN GEANT]{
        \label{fig:GEANT}
        \includegraphics[scale=0.22]{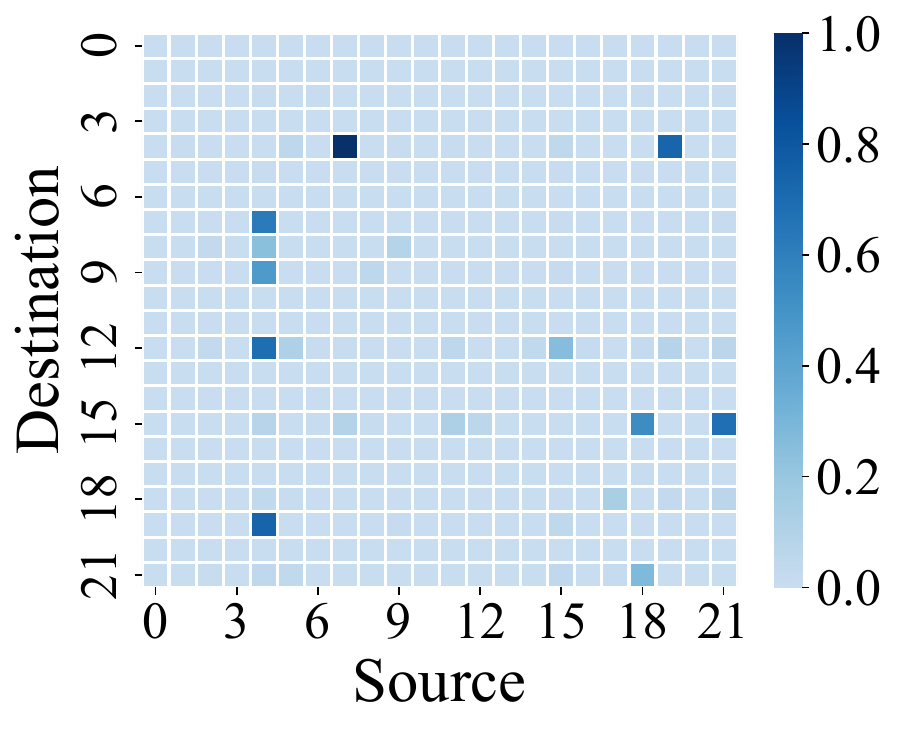}
        }
        \subfigure[PoD-level, Meta data center]{
        \label{fig:Facebook_pod_a}
        \includegraphics[scale=0.22]{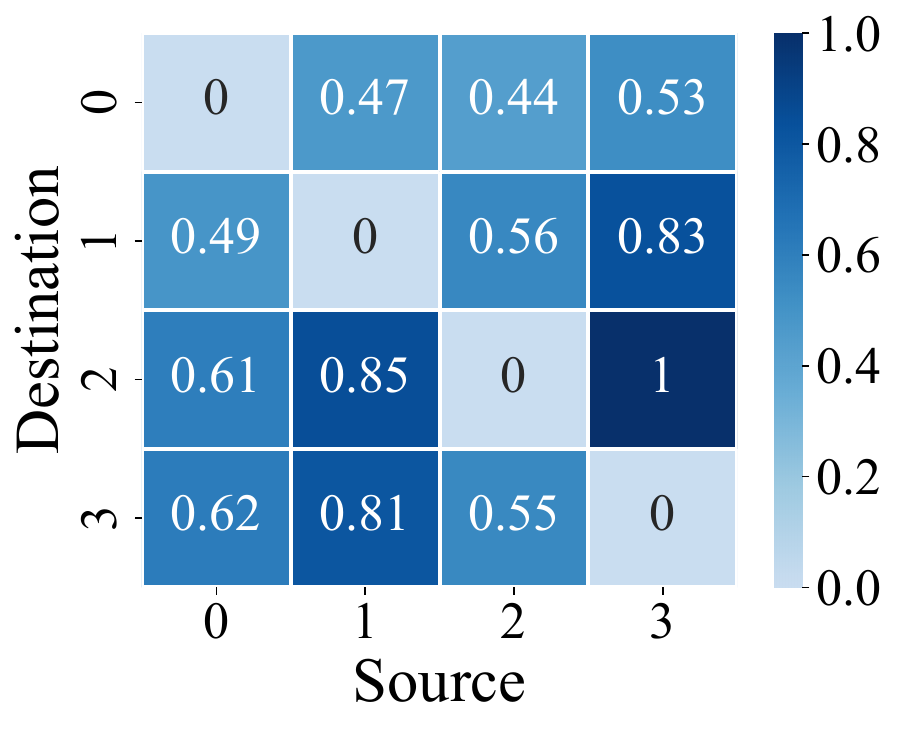}
        }
        \subfigure[ToR-level, Meta data center]{
        \label{fig:Facebook_tor_a}
        \includegraphics[scale=0.22]{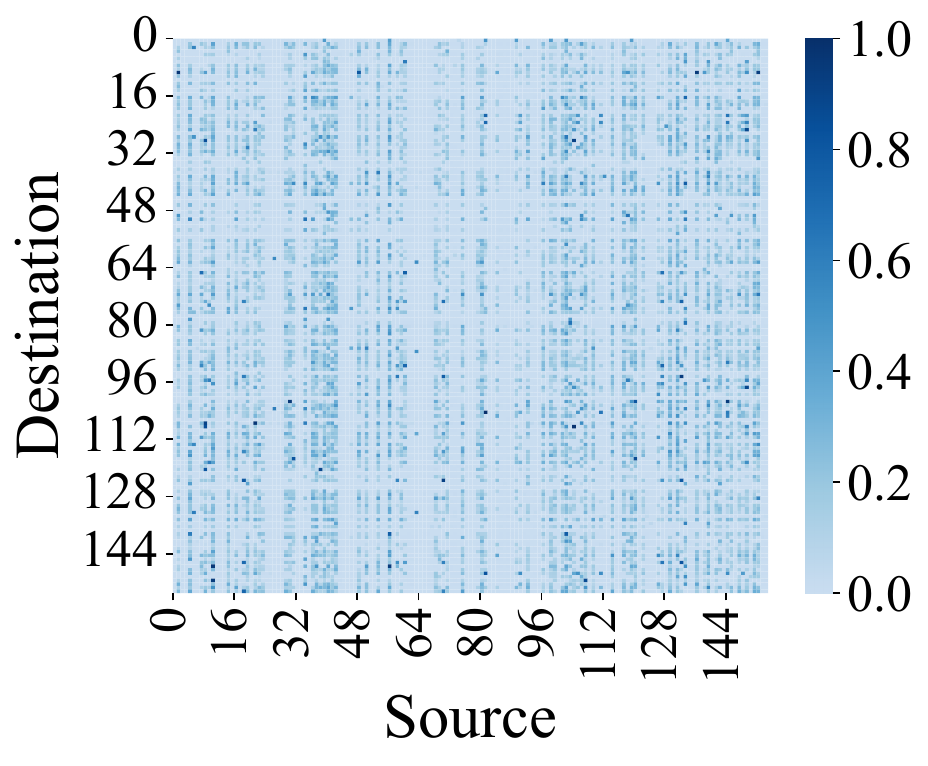}
        }
        \caption{The \textit{variance} of traffic demand by source and destination (The value of variance has been normalized). The larger the variance, the more likely it is that the traffic demand for that source-destination pair will experience a burst. Regardless of which network in \ref{fig:GEANT}, \ref{fig:Facebook_pod_a}, or \ref{fig:Facebook_tor_a}, the traffic demands of different source-destination pairs exhibit different traffic characteristics.}
        \label{fig:example of traffic demand}
    \end{minipage}%
    \hspace{3mm}
    \begin{minipage}{0.35\textwidth}
        \subfigure[Network traffic pattern.]{
        \label{fig:network traffic}
        \includegraphics[scale=0.23]{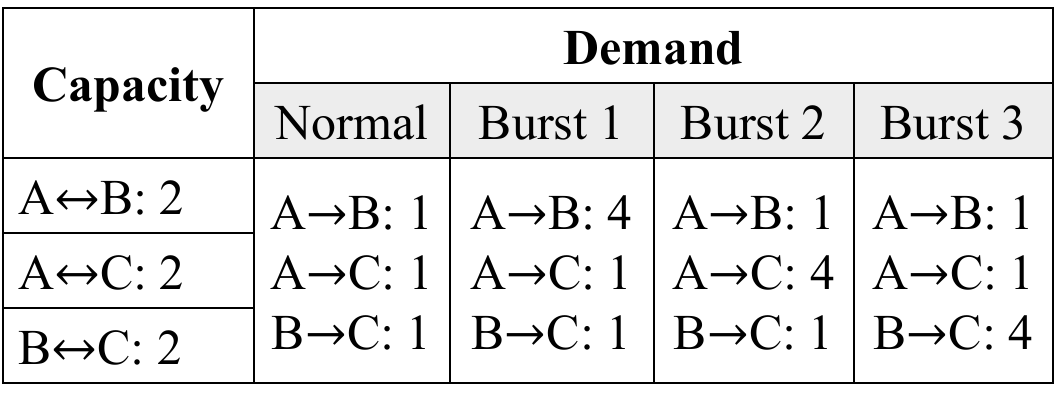}
        }
        \subfigure[Topology]{
        \label{fig:network topo}
        \includegraphics[scale=0.33]{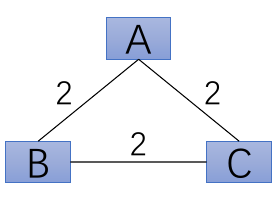}
        }
        \subfigure[TE scheme 1]{
        \label{fig:te scheme 1}
        \includegraphics[scale=0.375]{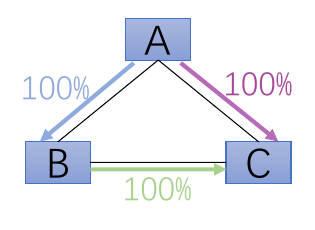}
        }
        \subfigure[TE scheme 2]{
        \label{fig:te scheme 2}
        \includegraphics[scale=0.375]{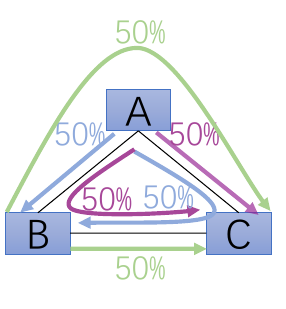}
        }
        \subfigure[TE scheme 3]{
        \label{fig:te scheme 3}
        \includegraphics[scale=0.375]{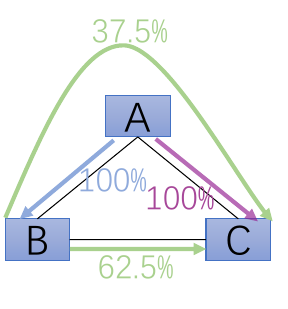}
        }
        \caption{A simple example comparing three TE schemes. TE scheme 1 has the optimal MLU in the normal situation. When traffic burst situations 1/2/3 are all possible, TE scheme 2 exhibits the best resilience against bursts. If only traffic burst situation 3 is likely to occur, TE scheme 3 outperforms TE scheme 2.}
        \label{fig:TE simple}
    \end{minipage}
\end{figure*}
To motivate the need to manage bursts in TE, we present an analysis of the impact of bursts on network performance and summarize the results in Figure \ref{fig: anti-burst motivation}. We implement two strategies in this study: 1) the `No hedging' strategy, which uses the current traffic matrix to determine the TE configuration for the next interval without any approach to manage bursts; 2) the `Hedging' strategy, which uses the current traffic matrix for configuring the next interval but incorporates the anti-burst Hedging mechanism utilized by Google in their Jupiter data center network \cite{poutievski2022jupiter}. The underlying principle of the Hedging mechanism is to spread a flow across multiple paths to prevent bursts from excessively impacting any single path. We conduct evaluations on the GEANT WAN \cite{uhlig2006providing}, and the PoD/ToR-level direct connect topologies\footnote{We change the topology from clos topologies to direct-connect topologies, because TE is rarely used in clos topologies. As optical circuit switching witnessed widespread adoption in Google's data centers~\cite{poutievski2022jupiter}, direct-connect topologies have become popular.} with traffic traces collected in Meta's data centers \cite{roy2015inside}. Since GEANT WAN only offers minute-level traffic matrices, for fair comparison, we also aggregate Meta's data center traffic trace at minute-level intervals. Our findings from Figure \ref{fig: anti-burst motivation} include:
\begin{itemize}[leftmargin=*]
    \item Performance sensitivity to network variability: From the GEANT WAN network to the PoD-level data center network, and then to the ToR-level data center network, traffic becomes more volatile, and the performance of the `No hedging' strategy becomes increasingly unstable.
    \item Necessity of burst resistance: Across WAN and data center networks at both the ToR and PoD levels, the `No hedging' strategy results in higher peaks on the MLU curve, indicating that the networks experience congestion due to bursts if anti-burst strategies are not employed.
    \item Performance trade-offs with burst resistance: The `No hedging' strategy displays higher peaks and lower troughs on the MLU curve. This can be interpreted as peaks occurring during bursts and troughs occurring in their absence. Conversely, the `Hedging' strategy does not achieve as low troughs during non-burst conditions, because it forces a significant portion of traffic to take non-optimal paths.
\end{itemize}

In summary, managing bursts is necessary for TE, yet strategies for burst management often compromise non-burst scenario performance. So we seek a TE method that effectively manages bursts while minimizing the impact on non-burst scenario performance.
\subsection{Diversity in traffic characteristics}
\label{section: diversity in traffic characteristics}
We find that although traffic bursts occur, the extent of these bursts varies among different source-destination pairs (SD pairs). To demonstrate the diverse traffic characteristics under different SD pairs, we provide an analysis of traffic characteristics across SD pairs in various production networks. This includes the WAN network GEANT \cite{uhlig2006providing}, as well as the Meta's data center network \cite{roy2015inside} at both the PoD-level and ToR-level. 
Details on the public data used are provided in \S \ref{section: evaluation}. For each SD pair, we calculate the variance of the traffic volumes as a measure of the dynamic changes in traffic. The results of our analysis are displayed in Figure \ref{fig:example of traffic demand}, which shows the \textit{variance} of traffic demand for different SD pairs across the three types of network traffic. The greater the variance for a SD pair, the more unstable that SD pair is. 

The results in Figure \ref{fig:example of traffic demand} indicate that, regardless of the topology type, the traffic characteristics are distinct for different SD pairs. If one treats the traffic from all the SD pairs equally, the resulting TE solution may either result in suboptimal performance in non-bursty scenarios or sacrifice the burst-handling capability. Hence, effectively leveraging the diverse traffic characteristics in TE is crucial for better balancing the performance trade-off between bursty and non-bursty situations.

\subsection{Dig deep into the trade-off}
\label{section: trade-off in te}
\textbf{Trade-off dilemma in TE.} 
To illustrate the trade-off dilemma in TE, we present an illustrative example in Figure \ref{fig:TE simple}. In this network, there are three traffic demands: $\text{A} \rightarrow \text{B}, \text{A} \rightarrow \text{C}, \text{B} \rightarrow \text{C}$. In the normal situation, the traffic demand for all three is $1$. However, in three different burst situations, the traffic demand for $\text{A} \rightarrow \text{B}, \text{A} \rightarrow \text{C}, \text{B} \rightarrow \text{C}$ increases to $4$, respectively. 

TE scheme 1 considers all traffic as non-bursty and only optimizes for the normal situation. To minimize congestion in non-bursty scenarios, TE scheme 1 directs all traffic along the shortest paths. In the normal situation, its Max Link Utilization (MLU) is $\max\{\frac{1}{2},\frac{1}{2},\frac{1}{2}\}=0.5$. When any burst situation occurs, the MLU is increased to $\max\{\frac{4}{2},\frac{1}{2},\frac{1}{2}\}=2$.

TE scheme 2 considers all traffic as bursty traffic and aims to enhance robustness in response to the three burst situations by splitting traffic across different paths. When dealing with the normal situation, the MLU of TE scheme 2 is $\max\{\frac{1\times0.5 +1\times0.5+1\times0.5}{2},\frac{1\times0.5 +1\times0.5+1\times0.5}{2},\frac{1\times0.5 +1\times0.5+1\times0.5}{2}\}=0.75$, while dealing with the three burst situations, its MLU is $\max\{\frac{4\times0.5 +1\times0.5+1\times0.5}{2},\frac{4\times0.5 +1\times0.5+1\times0.5}{2},\frac{4\times0.5 +1\times0.5+1\times0.5}{2}\}=1.5$. That is, TE scheme 2 exhibits better robustness in handling traffic bursts compared to TE scheme 1, but with a decrease in normal performance.

Overall, this one-size-fits-all approach either leads to a network lacking the capacity to handle bursts or enhances the capability to manage bursts but at the cost of compromised performance in non-burst situations.\\
\textbf{A key insight in balancing normal performance and robustness.}
We find that a more effective approach is to differentiate treatment based on the unique characteristics of each SD pair.
Figure \ref{fig:te scheme 3} illustrates this point. TE scheme 3 specifically addresses the bursts between $\text{B} \rightarrow \text{C}$, selecting two paths for serving the traffic from B to C. Conversely, for $\text{A} \rightarrow \text{B}$ and $\text{A} \rightarrow \text{C}$, it opts for direct paths. In TE scheme 3, the MLU is $\max\{\frac{1\times0.375 +1}{2},\frac{1\times0.375 +1}{2},\frac{1\times0.625}{2}\}=0.6875$ under the normal situation. When dealing with the traffic burst situation 1 or 2,
the MLU is $\max\{\frac{1\times0.375 +4}{2},\frac{1\times0.375 +1}{2},\frac{1\times0.625}{2}\}=2.1875$; when dealing with the traffic burst situation 3, the MLU is $\max\{\frac{4\times0.375 +1}{2},\frac{4\times0.375 +1}{2},\frac{4\times0.625}{2}\}=1.25$. It can be observed that, although TE scheme 3 is not as robust as TE scheme 2 in handling traffic burst situation 1/2, it performs better than TE scheme 2 in both normal situations and traffic burst situation 3. 
If the traffic demand from A to B and A to C never experiences traffic bursts, then TE scheme 3 would be a better solution compared to TE scheme 2. That is, if a traffic demand remains consistently stable, then the robustness of the paths serving that traffic demand does not need to be a concern (similar to how TE scheme 3 does not consider the robustness of path $\text{A} \rightarrow \text{B}$ and path $\text{A} \rightarrow \text{C}$).

Based on the observations in \S \ref{section: necessity of managing bursts} and the insights from Figure \ref{fig:example of traffic demand} and Figure \ref{fig:TE simple}, it is evident that enhancing robustness to manage unexpected traffic bursts is necessary. However, uniformly increasing the robustness across all paths may lead to a compromise in average performance. A more nuanced strategy would be to adjust the robustness of paths based on the traffic characteristics of the traffic demands they serve. 
Compared to previous TE methods,
this fine-grained robustness enhancement strategy in TE enables a better balance between normal-case performance and burst-case performance.
\section{TE model}
\label{secton:te model}
\textbf{Notations \& Definitions.} We introduce recurring mathematical notations and definitions of TE. All notations used in this paper are also tabulated in Table \ref{table: notations used in this paper} for ease of reference.
\begin{itemize}[leftmargin=*]
    \item \textbf{Network.} We represent the network topology as a graph $G=(V, E, c)$, where $V$ and $E$ are the vertex and edge sets, respectively, and $c:E \rightarrow \mathbb{R}^+$ assigns capacities to edges.
    \item \textbf{Traffic demands.} A Demand matrix (DM) $D$ is a $|V|\times|V|$ matrix whose $(i,j)$'th entry $D_{ij}$ specifies the traffic demand between source $i$ and destination $j$. 
    
    \item  \textbf{Network paths.} Each source vertex $s$ communicates with each destination vertex $d$ via a set of network paths $P_{sd}$. We can also say that $P_{sd}$ serves the SD pair $(s,d)$. 
    \item  \textbf{Path capacity.} The capacity of a path $p$ is denoted by $C_p$. It is the minimum capacity across the edges along the path \cite{poutievski2022jupiter}.
    \item \textbf{TE configuration.} A TE configuration $\mathcal{R}$ specifies for each source vertex $s$ and destination vertex $d$ how the $D_{sd}$ traffic from $s$ to $d$ is split across the paths in $P_{sd}$. A TE configuration specifies for each path $p \in P_{sd}$ a split ratio $r_{p}$, where $r_p$ is the fraction of the traffic demand from $s$ to $d$ forwarded along path $p$. The split ratios must satisfy $\sum_{p \in P_{sd}}r_p = 1$.
    \item \textbf{TE objective.} Our TE objective is to minimize the Max Link Utilization (MLU), which is a classical TE objective \cite{poutievski2022jupiter,azar2003optimal,valadarsky2017learning,chiesa2016lying,benson2011microte}. Google found MLU to be a reasonable proxy metric for throughput as well as for resilience against traffic pattern variation. They found high MLU indicates many links are in danger of overloading, causing packet losses, increasing flow-completion time, and reducing throughput \cite{poutievski2022jupiter}.

    Given a demand matrix $D$ and TE configuration $\mathcal{R}$, the total traffic traversing an edge $e$ is $f_e = \sum_{s,d \in V, p\in P_{sd},e\in p}D_{sd}\cdot r_p$, and then the MLU induced by $D$ and $\mathcal{R}$ is $\text{MLU} = \max_{e\in E}\frac{f_e}{c(e)}$, which we denote as $M(\mathcal{R},D)$. 
\end{itemize}
\textbf{TE under traffic uncertainty.} 
At epoch \( t \), the decision maker in TE must output a network configuration \( \mathcal{R}_t(D_1,\dots, D_{t-1}) \) based on historical data \(\{D_1,\dots, D_{t-1}\}\) before the arrival of \( D_t \). The objective is to obtain a network configuration that yields low link over-utilization for the upcoming traffic \( D_{t} \).
\begin{equation}
\label{equation: objective of decision maker}
\begin{aligned}
&\mathop{\text{minimize}}\limits_{\mathcal{R}_t(D_1,\dots,D_{t-1})}\qquad M(\mathcal{R}_t(D_1,\dots,D_{t-1}),D_t) 
\end{aligned}
\end{equation}
When configuring \( \mathcal{R}_t \) based on historical data \(\{D_1,\dots,D_{t-1}\}\) \textit{before} the arrival of $D_t$, it is critical to consider both anticipated and unanticipated traffic. An anticipated traffic demand \( D_t^{\text{expect}} \) can be deduced from historical traffic data using certain traffic prediction technologies. However, we caution that such a prediction may not be accurate due to the existence of highly bursty traffic demands. To attain better network performance, it is crucial to incorporate the mismatch between the predicted traffic demand \( D_t^{\text{expect}} \) and the real traffic demand $D_t$ when calculating each TE configuration.

\section{FIGRET design}
\label{section: TERMOL design}
In this section, we present the design of FIGRET. In \S \ref{section: design for managing mis-prediction}, we introduce how FIGRET addresses traffic bursts while solving TE solutions based on historical traffic. Subsequently, \S \ref{section: figret framework} discusses the transformation of problem-solving into the design of loss functions using deep learning. The design of the loss function and the deep neural network architecture are then detailed in \S \ref{section: design loss function of figret} and \S \ref{section: design dnn structure of figret}, respectively. Finally, \S \ref{section:coping with link failures} discusses how FIGRET handles node/link failures.
\subsection{Design for managing traffic bursts}
\label{section: design for managing mis-prediction}
As outlined in \S \ref{secton:te model}, it is not adequate for TE to rely solely on \( D^{\text{expect}} \) obtained from historical data. To augment robustness, the impact of traffic bursts must be considered. \\
\textbf{By defining a set to account for traffic bursts}. A prevalent method involves defining the bursts as a set \( \Delta \). During the optimization process, the sum of a specific burst $\delta \in\Delta$ and \( D^{\text{expect}} \) is used as an estimate for the actual traffic, denoted as \( \hat{D_t} = \delta + D^{\text{expect}} \). The objective is to find a TE configuration that offers the best worst-case guarantee for all the $\delta \in\Delta$, i.e.,
\begin{equation}
\boxed{
\begin{aligned}
&\mathop{\text{minimize}}\limits_{\mathcal{R}_t}\qquad \text{max}_{\hat{D_t}\in\hat{\mathcal{D}}} M(\mathcal{R}_t,\hat{D_t})\\
&\text{subject to}\qquad\hat{\mathcal{D}}=\{\hat{D_t} | \hat{D_t}=\delta + D^{\text{expect}},\delta\in\Delta\}
\end{aligned}
}
\label{equation: set burst}
\end{equation}

\textit{Drawback}: The above robust optimization based approach could offer varying levels of robustness with different choices of $\Delta$, but still faces the following shortcomings:

\begin{itemize}[leftmargin=*]
    \item It is hard to find the most appropriate $\Delta$ that yields the best trade-off between normal case performance and burst case performance. Although one could use Chebyshev's inequality~\cite{marshall1960multivariate} to quantify a $\Delta$ such that the probability $P(D_t\notin \hat{\mathcal{D}})$ is no larger than a small value $\alpha$, $D_t$ may still go beyond $\hat{\mathcal{D}}$ and the performance guarantee offered by Equation (\ref{equation: set burst}) becomes invalid. Admittedly, we could increase $\Delta$ such that the probability $\alpha$ is extremely small, but the normal-case performance could then be compromised.
    \item The objective function in Equation (\ref{equation: set burst}) aims to optimize the worst case within the set $\hat{\mathcal{D}}$. Note that $\hat{\mathcal{D}}$ may contain an infinite number of distinct traffic patterns, dramatically increasing the computational complexity of solving Equation (\ref{equation: set burst}). In contrast, our chosen anti-burst strategy in Equation (\ref{equation: te opt, robustness by constraint}) only requires optimization for a single traffic pattern $D_t^{\text{expect}}$.
\end{itemize}
\textbf{By imposing constraints to account for traffic bursts.} 
When considering both \( D^{\text{expect}} \) and \( \delta \), the utilization of each edge can be expressed as: \( \sum_{s,d \in V, p \in P_{sd}, e \in p} (D_{sd}^{\text{expect}} + \delta_{sd}) \cdot r_p / c(e) \). Consequently, the actual impact of \( \delta_{sd} \) on each edge can be represented by its coefficient. To mitigate the impact of traffic bursts, we could impose an upper bound on this coefficient. The degree of robustness can be controlled by adjusting the upper bound of each coefficient.

As discussed in \S \ref{section: motivation}, we aim to impose varying levels of robustness on different paths based on the traffic characteristics of the SD pairs. Therefore, this constraint should be a function related to the SD pairs, defined as \( \mathcal{F}:(s,d)\rightarrow \mathbb{R}^+ \).
\begin{equation}
    \forall s,d \in V, \forall p\in P_{sd},\forall e\in P \quad \frac{r_p}{c(e)}\leq \mathcal{F}((s,d)).
\label{equation: edge sensitivity constraint}
\end{equation}
Considering the definitions of path capacity $C_p$, as expressed by $C_p = \min_{e\in p}c(e)$, it follows that Equation (\ref{equation: edge sensitivity constraint}) is equivalent to Equation (\ref{equation: path sensitivity constraint}).
\begin{equation}
    \forall{s,d\in V}, \forall{p\in P_{st}} \quad \frac{r_p}{C_p} \leq \mathcal{F}((s,d)).
\label{equation: path sensitivity constraint}
\end{equation}
As indicated in Equation (\ref{equation: path sensitivity constraint}), ensuring that the ratio $\frac{r_p}{C_p}$ falls below a certain limit guarantees the robustness of network configurations against burst traffic. Therefore, $\frac{r_p}{C_p}$ can be used as a metric to represent the level of robustness in the network configuration. So we define path sensitivity $\mathscr{S}_p$ for path $p$ as
\begin{equation*}
    \mathscr{S}_p = \frac{r_p}{C_p}.
\end{equation*}
By taking into account the \emph{path sensitivity} metric in our TE formulation, the impact of traffic bursts can be mitigated. The resulting formulation is shown below:
\begin{equation}
\label{equation: te opt, robustness by constraint}
\boxed{
\begin{aligned}
&\mathop{\text{minimize}}\limits_{\mathcal{R}_t}\qquad M(\mathcal{R}_t,D_t^{\text{expect}})\\
&\text{subject to}\qquad \mathscr{S}_p \leq \mathcal{F}((s,d)), \forall s,d\in V, \forall p\in P_{sd}
\end{aligned}
}
\end{equation}
\textbf{Remark.} In Equation (\ref{equation: te opt, robustness by constraint}), both \(D_t^{\text{expect}}\) and \(\mathcal{F}\) are determined based on historical traffic data \(\{D_1, \dots, D_{t-1}\}\). Selecting an appropriate \(D_t^{\text{expect}}\) and \(\mathcal{F}\) for a specific network topology and traffic pattern is not trivial. Existing methods have not finely discriminated between different SD pairs. For example:
\begin{itemize}[leftmargin=*]
    \item Traditional methods based on direct optimization of MLU after prediction, as well as those based on deep learning \cite{valadarsky2017learning,perry2023dote}, set \( \mathcal{F}((s,t)) \equiv +\infty, \forall s,t \in V \). They assume that the real flow can be entirely represented by \( D^{\text{expect}}_t \). However, due to the inherent dynamic nature of traffic demands, mismatches are inevitable.
    \item Desensitization-based TE schemes \cite{poutievski2022jupiter} set \( \mathcal{F}((s,d)) \equiv \text{const}, \forall s,d \in V \). A single constant value makes it challenging to perfectly balance burst resistance and average performance. Another approach \cite{teh2022enabling} sets $\mathscr{S}_p$ as an objective, aiming to minimize the maximum $\mathscr{S}_p$ among all paths. However, it also does not account for the difference among different SD pairs.
\end{itemize}
\subsection{FIGRET framework}
\label{section: figret framework}
\subsubsection{How to compute a TE solution}
Given the FIGRET formulation in Equation (\ref{equation: te opt, robustness by constraint}), the next challenge is to efficiently compute TE solutions. Two approaches to consider are the two-stage method and the end-to-end method.

The \textit{two-stage method} first explicitly estimates $D_t^{\text{expect}}$ and $\mathcal{F}$. Then, it employs a linear programming-based algorithm to solve Equation (\ref{equation: te opt, robustness by constraint}). However, this approach has the following shortcomings, making it far from ideal:
\begin{itemize}[leftmargin=*]
    \item Due to the existence of highly bursty source-destination pairs, predicting a suitable $D_t^{\text{expect}}$ is challenging.
    \item There exists a mismatch between the upstream tasks of determining \(D_t^{\text{expect}}\) and \(\mathcal{F}\), and the downstream task of optimizing MLU. Predicting \(D_t^{\text{expect}}\) often relies on Mean Squared Error (MSE) for evaluation, which does not account for network topology—an essential factor in MLU optimization. For instance, in SD pairs with high-capacity paths, the precision of traffic predictions becomes less critical, as illustrated in Figure \ref{fig: mismatch in prediction} in Appendix \ref{section:objective mismatch}. This oversight can adversely affect TE performance. A similar mismatch affects \(\mathcal{F}\), where constraints should reflect not only traffic characteristics but also the topological structure.
    \item Employing linear programming involves high computational complexity and may not scale to large networks.
\end{itemize}

The \textit{end-to-end method} avoids the explicit prediction of $D_t^{\text{expect}}$ and $\mathcal{F}$. Instead, it directly establishes a relationship between historical traffic data \(\{D_1, \dots, D_{t-1}\}\) and a TE configuration $R_t$, with a goal to minimize MLU and ensure robustness. This method’s advantage lies in its ability to consider the impact of network topology, eliminating the mismatch between the upstream traffic prediction and the downstream network optimization \cite{perry2023dote}. Thus, its performance is better than that of the two-stage method (See Appendix \ref{section: heuristic selection} for more details).

Overall, based on the above considerations, we opt for the end-to-end method to compute TE solutions.
\subsubsection{Leveraging deep learning to implement the end-to-end method}
We employ an end-to-end method to establish a mapping between historical traffic data \(\{D_1, \ldots, D_{t-1}\}\) and TE configuration \(R_t\) that minimizes MLU while ensuring robustness. Since this end-to-end method does not pre-solve $D_t^{\text{expect}}$, we are unable to establish a clear optimization problem as in Equation (\ref{equation: te opt, robustness by constraint}). Therefore, we cannot utilize a linear programming-based approach to implement the end-to-end method. In response, drawing inspiration from DOTE \cite{perry2023dote}, FIGRET adopts a deep neural network (DNN) to implement the end-to-end method. By designing a well-designed burst-aware loss function, FIGRET can output TE solutions that effectively balance normal and burst traffic scenarios.
\subsection{Design loss function of FIGRET}
\label{section: design loss function of figret}
In FIGRET, a Deep Neural Network (DNN) is employed to establish a mapping between historical traffic data \(\{D_{1}, \dots, D_{t-1}\}\) and TE configurations. To maintain a consistent input size for the DNN and prevent an overload of traffic data when \(t\) is large, a temporal window \(H\) is typically chosen \cite{valadarsky2017learning,perry2023dote}. During training, FIGRET receives \(\{D_{t-H}, \dots, D_{t-1}\}\), and subsequently outputs a TE Configuration \(R_t\). We denote this output as \(R_t = \pi_\theta(D_{t-H}, \dots, D_{t-1})\), where \(\pi\) represents the mapping function from DNN inputs to outputs, and \(\theta\) denotes the DNN's link weights. The revealed \(D_t\) allows for the calculation of the loss for this \(R_t\) through the loss function \(\mathcal{L}(R_t, D_t)\), after which a gradient descent optimizer updates the parameters \(\theta\). Thus, to enable FIGRET to output TE solutions with fine-grained robustness, a \textit{well-designed loss function} is essential. 

The design of our loss function is intrinsically aligned with the Lagrangian relaxation of Equation (\ref{equation: te opt, robustness by constraint}), as delineated in Equation (\ref{equation: lagrangian_relaxation}). This relaxation framework guides us in structuring a loss function that encompasses two components.
\begin{equation}
\label{equation: lagrangian_relaxation}
\boxed{
\begin{aligned}
&\mathop{\text{minimize}}\limits_{\mathcal{R}_t, \lambda}\quad L(\mathcal{R}_t, \lambda) = M(\mathcal{R}_t, D_t^{\text{expect}}) \\
&\quad + \sum_{s,d \in V} \sum_{p \in P_{sd}} \lambda_{sd} \left( \mathscr{S}_p - \mathcal{F}((s,d)) \right) \\
&\text{subject to}\quad \lambda_{sd} \geq 0, \quad \forall s,d \in V, \forall p \in P_{sd}
\end{aligned}
}
\end{equation}
\subsubsection{Loss for MLU}
 The first component corresponds to minimizing max link utilization.
\begin{equation}
\mathcal{L}_1 = M(R_t,D_t).
\end{equation}

With the first loss function in place, upon completion of training, FIGRET \textit{implicitly} learns a probability distribution \(P(D_t^{\text{expect}}|D_{t-1},\dots,D_{t-H})\). It can then output an \(R_t\) that optimizes \(\mathbb{E}_{D_t^{\text{expect}}}[M(\mathcal{R}^t, D_t^{\text{expect}})]\) with respect to \(P(D_t^{\text{expect}}|D_{t-1},\dots,D_{t-H})\).

\subsubsection{Loss for fine-grained robustness}
The second component of the loss function reflects the constraints imposed by path sensitivities. Since we adopt an end-to-end TE approach without explicitly solving for $\mathcal{F}((s,d))$ beforehand, we can not directly use the second term of Equation (\ref{equation: lagrangian_relaxation})—\(\sum_{s,d \in V} \sum_{p \in P_{sd}} \lambda_{sd} (\mathscr{S}_p - \mathcal{F}((s,d)))\)—as the loss function. In response, we design the following heuristic scheme for constructing the second component of the loss.

\textbf{How to calculate $\mathcal{L}_2$?} We use \( \sigma^2_{D_{sd},[1-T]} \) to denote the variance of traffic demands \( D_{sd} \) from source \( s \) to destination \( d \) over the training period from time 1 to \( T \). We denote \( \mathscr{S}_{sd}^{\text{max}} \) as the maximum path sensitivity among all paths serving from source \( s \) to destination \( d \) in the configuration $\mathcal{R}_t$.
 Based on this, we can express $\mathcal{L}_2$ as Equation (\ref{equation:loss}).

\begin{equation}
\mathcal{L}_2 = \sum_{\forall s,d \in V} \sigma^2_{D_{sd},[1-T]}\times \mathscr{S}_{sd}^{\text{max}}.
\label{equation:loss}
\end{equation}

\textbf{How does Equation (\ref{equation:loss}) capture fine-grained robustness?} 
The original term \(\sum_{s,d \in V} \sum_{p \in P_{sd}} \lambda_{sd} (\mathscr{S}p - \mathcal{F}((s,d)))\) facilitates fine-grained robustness because the values of $\mathcal{F}((s,d))$ differ among various source-destination pairs, resulting in different penalties for the same path sensitivity across different paths. To elaborate, for SD pairs with stable traffic patterns, denoted as \((s,d)_{\text{stable}}\), and those characterized by bursty traffic dynamics, labeled as \((s,d)_{\text{bursty}}\), the constraints are more stringent for bursty SD pairs, i.e., \(\mathcal{F}((s,d)_{\text{stable}}) > \mathcal{F}((s,d)_{\text{bursty}})\). Therefore, according to $\mathscr{S}p - \mathcal{F}((s,d)))$, the path serving for bursty traffic SD pair incurs a greater penalty.
Our design of $\mathcal{L}_2$ achieves a similar functionality. We calculate the variance of traffic demand \(\sigma^2_{D_{sd},[1-T]}\) for each SD pair from time 1 to \(T\), representing the degree of variation in the flow. This provides a quantitative measure of traffic fluctuation for the network. Subsequently, we weight these variances with the maximum path sensitivity \(\mathscr{S}_{sd}^{\text{max}}\) of paths serving the corresponding SD pair. This approach imposes stricter sensitivity penalties on flows with higher variance and lighter penalties on those with lower variance. 

When training the neural network, we consider both the loss term MLU \( \mathcal{L}_1 \) and the robustness enhancement term \( \mathcal{L}_2 \). This comprehensive consideration aims to guide the network in learning how to adjust path sensitivity constraints for different SD pairs while maintaining MLU optimization. By this method, we ensure that the network not only meets the MLU optimization objectives but also enhances the burst resistance of different SD pairs in a fine-grained manner.
\subsection{Design DNN structure of FIGRET}
\label{section: design dnn structure of figret}
The structures primarily used in TE are as follows: Graph Neural Network (GNN) \cite{xu2023teal, bernardez2021machine}, Convolutional Neural Network (CNN) \cite{valadarsky2017learning}, and Fully Connected Network (FCN) \cite{perry2023dote}. In this section, we briefly explain why we chose FCN over GNN and CNN, with more detail available in Appendix \ref{section: dnn architecture details}.\\
\textbf{Network topology handling: no need for GNN}. In TE problems, acquiring and utilizing network topology information is crucial for constructing the mapping between TE configurations and MLU. The network topology \( G(V, E, c) \), as a graph structure, seems naturally suited for processing with GNN. However, as Function \ref{algorithm: mapping MLU} in Appendix \ref{section: no need for GNN} shows, this mapping relationship can be achieved through simple matrix operations, which can be fully handled by FCN. GNN involves processing based on adjacency matrices and node features, which can lead to redundant computations and higher computational complexity, potentially consuming significant memory \cite{wang2019dynamic, ying2018graph, zhang2022understanding}. Considering these factors, using FCN seems a more reasonable choice.\\
\textbf{CNN vs. FCN: The inappropriateness of CNN.} CNN can effectively extract local information in data such as images through convolutional operations. However, in the historical traffic demand data used as input for TE, there is no obvious local information. The traffic sharing a common edge can be spatially close or distant, not necessarily confined to a local area. A detailed example is provided in Figure \ref{fig: cnn}. Therefore, we opted not to use CNN, but rather FCN.

\subsection{Coping with node/link failures}
\label{section:coping with link failures}
In large networks, node/link failures are inevitable, causing some paths unavailable to serve traffic.
A widely-adopted approach to addressing network failures in TE involves rerouting traffic around failed paths by allowing traffic sources to proportionally redistribute traffic among their remaining paths \cite{hong2013achieving, jain2013b4, suchara2011network, perry2023dote}.
\begin{itemize}[leftmargin=*]
\item In cases where the remaining paths have allocation ratios, the traffic demand from a failed path is proportionally redistributed based on these ratios. Consider a scenario with three paths having allocation ratios of (0.5, 0.3, 0.2). If the first path experiences a failure, the adjusted distribution ratios for the remaining paths would be set as (0, 0.6, 0.4).
\item Conversely, if the remaining paths do not have allocation ratios, the traffic demand from the failed path is equally distributed among them. For instance, in a situation with three paths at ratios (1, 0, 0), and the first path fails, the redistributed traffic is allocated evenly, resulting in new ratios of (0, 0.5, 0.5).
\end{itemize}
We integrate this approach into FIGRET, evaluate its effectiveness, and demonstrate that it achieves high resiliency to failures in \S \ref{section: coping with network failures}. Notably, handling link failures in FIGRET does not necessitate retraining.

\section{evaluation}
\label{section: evaluation}
In this section, we evaluate FIGRET. In \S \ref{section: methodology}, we first describe our evaluation methodology and analyze the types of traffic used in the evaluation to demonstrate that our experiments cover a wide variety of traffic scenarios. Subsequently, in \S \ref{section: comparing figret with other TE schemes}, we compare FIGRET with state-of-the-art TE schemes, focusing on TE quality, solution times, and precomputation times. Then, the effectiveness of FIGRET in handling link failures and adapting to drifts in traffic patterns is evaluated in \S \ref{section: coping with network failures} and \S \ref{section:robustness to demand changes}, respectively. Finally, in \S \ref{section: interpretation of termol}, we showcase the fine-grained robustness enhancement capabilities of FIGRET, thereby providing an interpretation of FIGRET's performance.
\subsection{Methodology}
\label{section: methodology}
\textbf{Topologies.} We consider three WAN topologies, including the pan-European research network, GEANT \cite{uhlig2006providing}, as well as two topologies from the Internet Topology Zoo \cite{knight2011internet}, namely UsCarrier and Kdl. We also examine various data center topologies. This includes the pFabric \cite{alizadeh2013pfabric}, which utilizes a leaf-spine topology with nine Top-of-Rack (TOR) switches. We change the topology from leaf-spine to a fully connected direct-connect network, because TE is rarely used in leaf-spine topologies. Additionally, we consider two Meta's DC clusters \cite{roy2015inside}: the Meta DB cluster and the Meta WEB cluster. The Meta DB cluster comprises MySQL servers that store user data and handle SQL queries, whereas the Meta WEB cluster is responsible for serving web traffic. For each cluster, we contemplate two types of topologies: the Top-of-Rack (ToR) level and the Point of Delivery (PoD) level. For Meta's data centers, we also adopt the direct-connect network as the topology. For the PoD level topology, we consider the fully connected network, whereas, for the ToR level topology, we utilize the random regular graph \cite{singla2012jellyfish}. Their numbers of nodes and edges are summarized in Table \ref{table: topologies}. For each topology, unless otherwise stated, we employ Yen’s algorithm to precompute the three shortest paths between every pair of nodes as the candidate paths for flow allocation \cite{abuzaid2021contracting,narayanan2021solving}. 
\begin{table}[ht]
\centering
\scalebox{0.85}{
\begin{tabular}{cccc}
\hline
                              & \textbf{\#Type} & \textbf{\#Nodes} & \textbf{\#Edges} \\ \hline
GEANT                         & WAN             & 23               & 74               \\
UsCarrier                     & WAN             & 158              & 378              \\
Cogentco                      & WAN             & 197              & 486              \\ \hline
pFabric                       & ToR-level DC    & 9                & 72               \\ \hline
\multirow{2}{*}{Meta DB}  & PoD-level DC    & 4                & 12               \\
                              & ToR-level DC    & 155              & 7194             \\ \hline
\multirow{2}{*}{Meta WEB} & PoD-level DC    & 8                & 56               \\
                              & ToR-level DC    & 324              & 31520            \\ \hline
\end{tabular}
}
\caption{Network topologies in our evaluation.}
\label{table: topologies}
\end{table}

\begin{figure}[!h]
    \centering
    \includegraphics[scale=0.33]{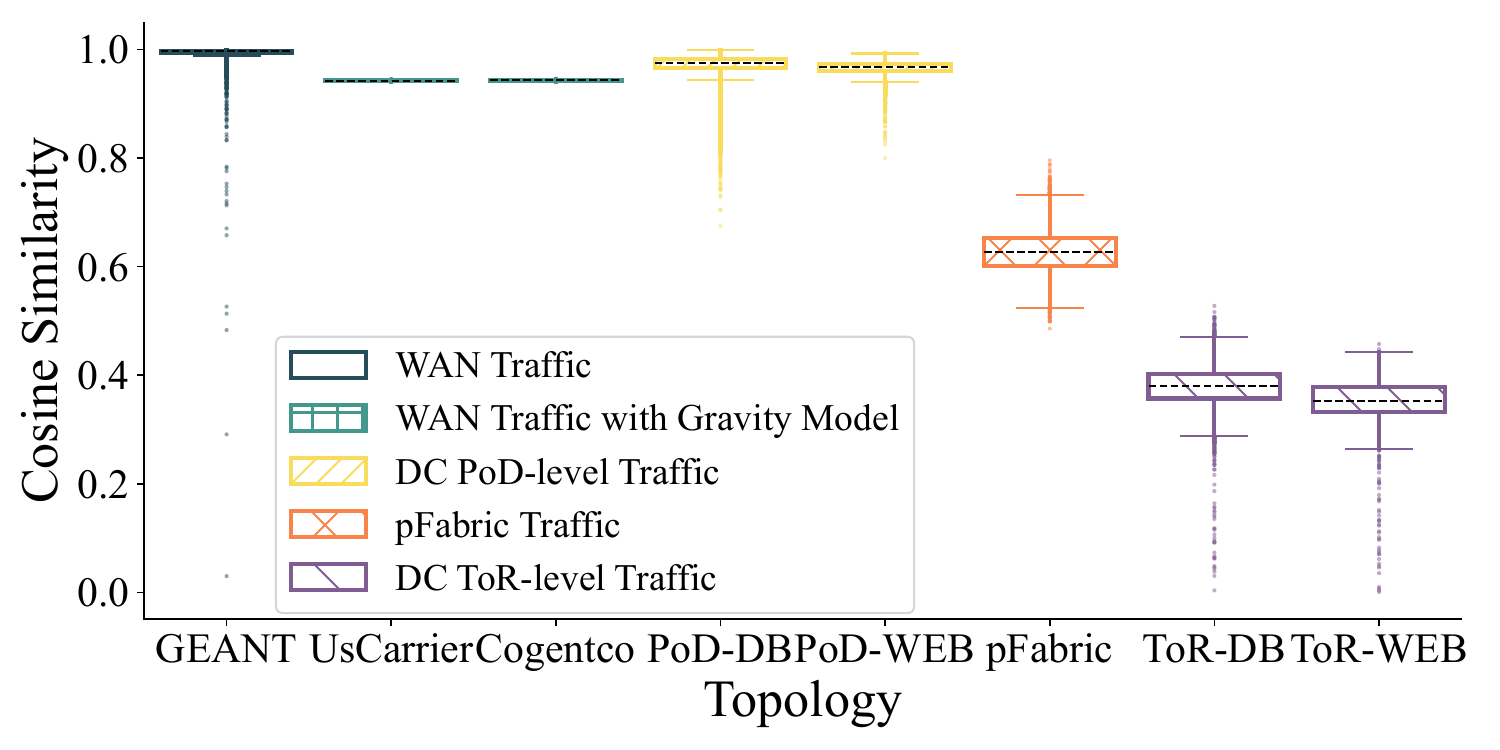}
    \caption{Cosine similarity analysis using a window of 12 historical TMs vs. the currently-seen TM.}
    \label{fig:Cosine similarity}
\end{figure}

\noindent\textbf{Traffic data.} For the GEANT topology, there are publicly available traffic matrix traces, which consist of data aggregated every 15 minutes over four months \cite{uhlig2006providing}. In the context of pFabric topology, the pFabric trace is characterized by a Poisson arrival process. When a flow arrives, the source and destination nodes are chosen uniformly at random from the different ToR switches. The size of each flow is determined randomly, adhering to the distribution outlined in the ``web search workload'' scenario as described in \cite{alizadeh2013pfabric}. In the case of the Meta topology, there is a public trace of one day's traffic \cite{roy2015inside}.
For the PoD-level topology, we aggregate traces into 1-second snapshots of the inter-Pod traffic matrix, while for the ToR-level topology, aggregation is performed into 10-second snapshots of the inter-ToR traffic matrix. The longer aggregation time for ToR-level topologies is attributed to their larger scale, which involves more nodes and a more complex network structure, resulting in greater computational and storage demands for TE.
For the UsCarrier and Cogentco topologies from Topology Zoo, where no public traffic traces are available, we employ a gravity model \cite{applegate2003making,roughan2002experience} to generate synthetic traffic.\\
\textbf{Traffic characteristics.} Our data includes different types of topologies and their corresponding traffic, each with distinct traffic characteristics. By comparing the similarity of the currently-seen TM with historical TMs, we observe whether the traffic is relatively stable and predictable. Specifically, for every currently-seen TM, we consider a window of historical TMs, find the TMs that most closely resemble this currently-seen TM, and calculate the cosine similarity between the two. Figure \ref{fig:Cosine similarity} shows the distribution of the results. Each candlestick displays the distribution of cosine similarities for different traffics, with the box range extending from the 25th to the 75th percentile. The closer the distribution of a traffic's cosine similarity is to 1, the more similar it is to past traffic, indicating greater stability. Conversely, a cosine similarity closer to 0 suggests that the traffic is more erratic. 
We can identify the following characteristics:
\begin{itemize}[leftmargin=*]
    \item The degree of burstiness in traffic increases progressively from WAN traffic to DC POD-level traffic, to DC ToR-level traffic. This may be related to the aggregation of traffic; POD-level traffic aggregates ToR-level traffic, and the WAN connects different data centers, making the aggregation effect more pronounced. The aggregation of multiple traffic streams can cause the bursty variations of different traffic to `neutralize' each other, hence the more aggregation there is, the more stable the traffic becomes.
    \item Even though WAN traffic is generally stable, there are some anomalies in the cosine similarity distribution of publicly available WAN traffic, indicating that WAN traffic can sometimes experience sudden bursts.
    \item The WAN traffic generated by the gravity model resembles the trend of real WAN traffic. However, it cannot capture all the characteristics of real traffic, as it fails to reflect traffic bursts. In our experiments, we use it to understand the TE performance in stable traffic conditions.
\end{itemize}
In summary, the traffic data exhibits various characteristics, including profiles with low, moderate, and high burstiness. This allows for a thorough evaluation of the TE performance in various scenarios.

\noindent\textbf{Baseline.} We select the following TE schemes to compare with FIGRET:
(1) \textbf{Omniscient TE}: This scheme represents an optimization with perfect knowledge of future demands, providing a benchmark for the most efficient performance achievable in all TE schemes.
(2) \textbf{Desensitization-based TE}: This is the TE scheme used by Google in Jupiter data center\cite{poutievski2022jupiter} and \cite{teh2020couder}. This scheme constructs an anticipated matrix composed of the peak values for each source-destination pair within a time window. Then it optimizes the TE objective under the constraint that path sensitivity remains below a predetermined threshold.
(3) \textbf{Demand-oblivious TE} \cite{applegate2003making}: This scheme focuses on optimizing the worst-case performance across all possible traffic demands.
(4) \textbf{Demand-prediction-based TE} \cite{abuzaid2021contracting,hong2013achieving,jain2013b4}: This method involves predicting the next incoming traffic demand and configuring accordingly, \textit{without} considering the mis-predictions that may arise from the traffic uncertainty.
(5) \textbf{COPE} \cite{wang2006cope}: This scheme enhances demand-oblivious TE by also optimizing over a set of predicted traffic demands. It optimizes MLU across a set of DMs predicted based on previously observed DMs while retaining a worst-case performance guarantee.
(6) \textbf{Deep Learning-based TE (DOTE)} \cite{perry2023dote}: DOTE employs a DNN to directly output the TE configuration based on the traffic demand observed within a given time window.
(7) \textbf{TEAL} \cite{xu2023teal}: TEAL employs a combination of GNN and Reinforcement Learning (RL) to process a given traffic demand, subsequently outputting a network configuration tailored for this demand. In our experiments, due to the absence of prior knowledge about future traffic, we apply the TE solution computed from the traffic demand of the preceding time snapshot to the next time snapshot.
(8) \textbf{SMORE} \cite{kumar2018semi}: SMORE employs Räcke's oblivious routing paths \cite{racke2002minimizing} for path selection, with traffic splitting ratios optimized for the predicted future traffic demands.\\
\textbf{Infrastructure and software.} Our experiments are carried out on an Intel(R) Xeon(R) Silver 4110 CPU, equipped with 128GB of memory. Furthermore, Nvidia-Tesla P100 GPUs are accessible for all schemes, but only FIGRET, DOTE, and TEAL can utilize them. We implement FIGRET using PyTorch \cite{paszke2019pytorch}. For detailed information on neural network architecture, parameters, and optimizer settings, please refer to Appendix \ref{section: dnn architecture details}. For TE schemes that necessitate solving optimization problems, Gurobi (version 9.5.2) \cite{gurobi2021gurobi} is employed.
\subsection{Comparing with other TE schemes}
\label{section: comparing figret with other TE schemes}
\begin{figure*}[!h]
    \subfigure[GEANT $\&$ pFabric]{
    \label{fig: GEANT pFabric}
    \includegraphics[scale=0.55]{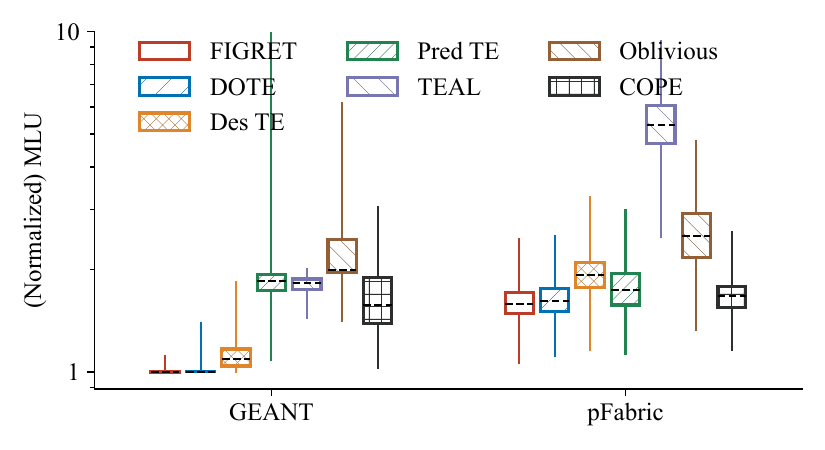}
    }
    \subfigure[ToR-level Meta DB $\&$ ToR-level Meta WEB]{
    \label{fig: tor_a tor_b}
    \includegraphics[scale=0.55]{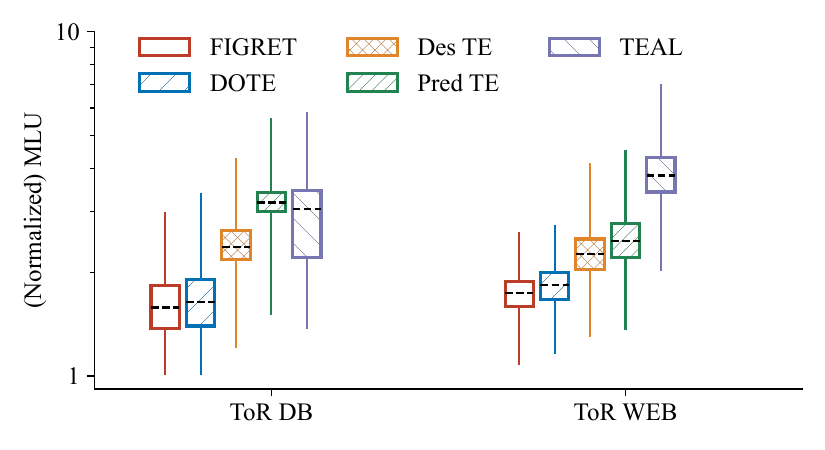}
    }
    \subfigure[PoD-level Meta DB $\&$ PoD-level Meta WEB]{
    \label{fig: pod_a pod_b}
    \includegraphics[scale=0.55]{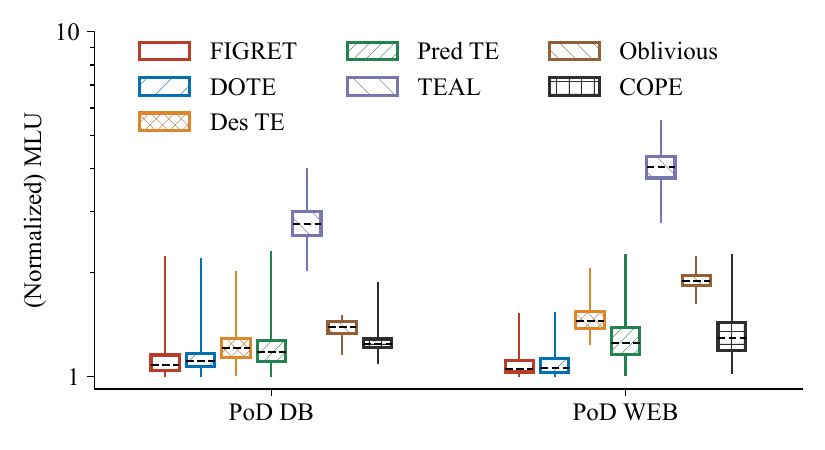}
    }
    \subfigure[Cogentco $\&$ UsCarrier]{
    \label{fig: zoo}
    \includegraphics[scale=0.55]{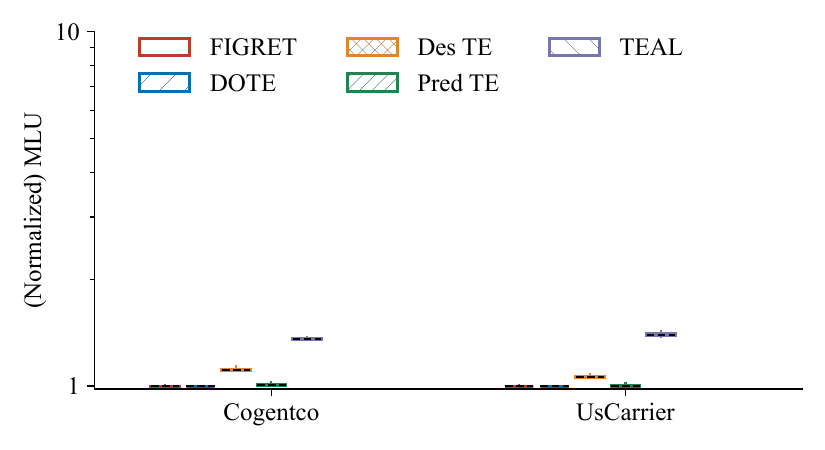}
    }
    \caption{Performance of FIGRET and baselines under the objective of minimizing max link utilization (MLU). The y-axis value represents the MLU normalized by that of the Omniscient TE.}
    \label{fig: performance of figret}
\end{figure*}
\textbf{TE quality.} Figure \ref{fig: performance of figret} compares the quality of TE solutions between FIGRET and other TE schemes. The y-axis value represents the MLU normalized by that of the Omniscient TE. We note the findings:
\begin{itemize}[leftmargin=*]
    \item \textbf{Desensitization-based TE.} Across all topologies, the performance of Desensitization-based TE in terms of both normal-case MLU (represented by the box) and the worst-case MLU (indicated by the top of the upper dashed line) is not satisfactory. This is due to the fact that Desensitization-based TE imposes unnecessary path sensitivity constraints on non-bursty traffic while providing insufficient path sensitivity constraints for bursty traffic. We will more clearly illustrate this using Figure \ref{fig: interpret termol} in \S \ref{section: interpretation of termol}.
    \item \textbf{Demand-prediction-based TE.} Demand-prediction-based TE methods primarily excel in environments with stable traffic demands. However, their effectiveness significantly diminishes when unexpected traffic bursts occur. The performance of these methods is highly dependent on the accuracy of traffic demand predictions. 
    \item \textbf{Demand-oblivious TE \& COPE.} Due to the extensive computation time and memory requirements of Demand-oblivious TE and COPE, we can only evaluate a few smaller-scale topologies. The two methods for handling mis-prediction are essentially the use of a set-based approach as introduced in \S \ref{section: design for managing mis-prediction}. It is evident that selecting a set not only increases computational complexity but also makes it difficult to achieve a good balance between normal-case performance and robustness to traffic bursts.
    \item \textbf{DOTE.} DOTE, the best-performing algorithm in existing TE \cite{perry2023dote}, leverages deep learning to map historical traffic data with the next network configurations, achieving outstanding normal-case performance. DOTE implicitly learns to achieve the lowest expected MLU with \( D^{\text{expect}} \), reaching the optimal average performance among existing methods. However, its drawback lies in handling burst situations. For example, in the GEANT data, as shown in Figure \ref{fig:Cosine similarity}, while most of the data points have high cosine similarity, close to 1, there are some outliers where the traffic's cosine similarity with past window traffic is very low, representing the unexpected burst scenarios cases. This leads to DOTE's MLU performance on GEANT data being close to that of Omniscient in most situations but with some peak values. As illustrated in Figure \ref{fig: tor_a tor_b}, in ToR-level DC with high dynamic characteristics, DOTE performs worse than FiGRET in terms of both average performance and robustness.
    \item \textbf{TEAL.} TEAL is designed to train a fast and scalable TE scheme. It receives a traffic demand and outputs a network configuration specifically for this demand. However, when this network configuration is applied to traffic demands with unexpected bursts, it tends to underperform. 
    \item  \textbf{FIGRET.} Our designed FIGRET achieves better results than previous TE methods, striking a better balance between normal-case performance and burst-case performance. For instance, compared to the TE scheme currently in Google's production data center network, FIGRET reduces the average MLU by 9\%-34\%. Against the state-of-the-art DOTE algorithm, FIGRET achieves a reduction in average MLU of 4.5\% and 5.3\% on ToR-level Meta DB and ToR-level Meta WEB, respectively, both characterized by high traffic dynamics. Even in stable topologies like WAN or Pod-level Datacenters, FIGRET performs no worse than DOTE. Simultaneously, we consider situations where the normalized MLU is greater than 2 as instances of severe congestion caused by inadequate network configuration. For the ToR-level Meta DB, FIGRET exhibits a 41\% lower incidence of severe congestion compared to DOTE, while for the ToR-level Meta WEB, the reduction is 53.9\%.
\end{itemize}
Furthermore, as shown in Figure \ref{fig: zoo}, all TE methods exhibit a low normalized MLU with no peaks. This is because the synthetic traffic generated by the gravity model is very stable and lacks traffic bursts. That said, handling traffic bursts is still critical for real WAN traffic because real WAN traffic does contain unexpected bursts. 

\begin{table*}
\centering
\scalebox{0.87}{
\begin{tabular}{c|cccc|cccc}
\hline
\multirow{2}{*}{Network}            & \multicolumn{4}{c|}{Calculation time (s)} & \multicolumn{4}{c}{Precomp. time (s)} \\
                                    & FIGRET  & LP   & Des TE & Oblivious \& Cope & FIGRET  & TEAL   & Oblivious  & COPE  \\ \hline
GEANT (\#Nodes 23, \#Edges 74)      & 0.002   & 0.04 & 0.07   & Feasible        & 150     & 1500   & 100        & 1000  \\
ToR DB (\#Node 155, \#Edges 7194)   & 0.005   & 1.60 & 5.00   & Infeasible      & 500     & 7000   & -          & -     \\
ToR WEB (\#Node 324, \#Edges 31520) & 0.009   & 7.30 & 17.00  & Infeasible      & 1500    & 20000  & -          & -     \\ \hline
\end{tabular}
}
\caption{Comparing the calculation and precomputation time across various TE schemes.}
\label{table: computing and precomputation}
\end{table*}

\noindent\textbf{Solver time.}
The solver time consists of two parts: the calculation time required to compute a new TE scheme for each new demand matrix, and the precomputation time.
Table \ref{table: computing and precomputation} presents a comparison of the calculation time for different TE schemes. Table \ref{table: computing and precomputation} lists four items: 1) FIGRET, 2) a Linear Programming (LP) approach that optimizes solely for expected traffic demands without applying any anti-burst strategy, 3) Desensitization-based TE, which handles traffic bursts by limiting path sensitivity, and 4) Oblivious \& COPE, which precompute TE solutions but do not update them thereafter; hence, we only test whether they can solve a problem instance. Clearly, FIGRET has the smallest calculation time, while Oblivious \& COPE TE schemes have the poorest scalability, failing to compute a solution for larger networks. The comparison between LP and Des TE shows that adding the path sensitivity constraints to handle traffic bursts increases the LP solver calculation time. In contrast, thanks to the adoption of deep neural network, FIGRET is able to enhance robustness without increasing solver complexity. Compared to Des TE, FIGRET achieves a speed-up of \(35\times\) to \(1800\times\).


Regarding the training time, since we employ a simple FCN architecture in FIGRET, the precomputation time required is shorter than COPE and reinforcement learning-based TE, especially when dealing with large-scale topologies. COPE and Demand-oblivious require substantial precomputation latency and a significant amount of memory. Given a maximum runtime of 1 day and machine memory constraints, they only completed the topologies for GEANT, pFabric, and Pod-level DC. Past public results also show their applicability is limited to smaller topologies \cite{wang2006cope,perry2023dote}. Simultaneously, compared to TEAL, which employs reinforcement learning and graph neural networks, FIGRET also has advantages in terms of training time. Detailed data are presented in Table \ref{table: computing and precomputation}.

\begin{figure}[!h]
    \centering
    \includegraphics[scale=0.5]{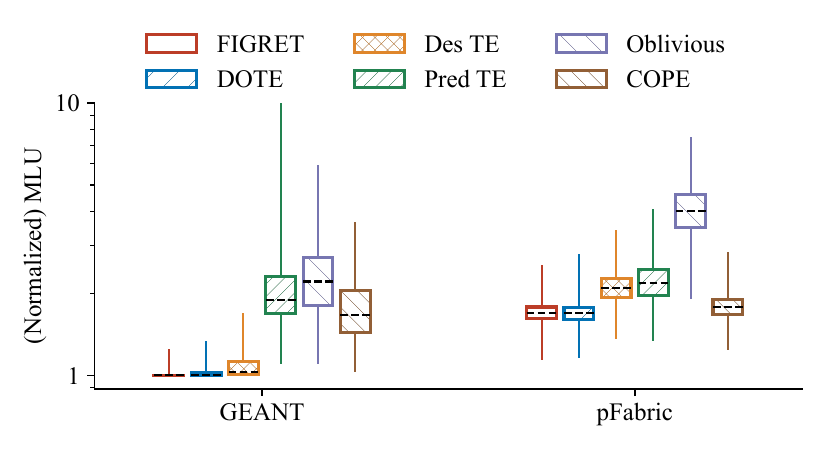}
    \caption{Performance of FIGRET and baselines with routing paths chosen by SMORE. ``Pred TE'' represents SMORE.}
    \label{fig:GEANT_and_pFabric_smore}
\end{figure}

\noindent\textbf{Path selection.} We compare FIGRET with SMORE. SMORE uses a set of paths computed by Räcke's oblivious routing algorithm. To ensure fairness, we use the same Räcke’s routing algorithm to generate paths for other algorithms. In this case, the prediction-based TE coincides with SMORE. The results are summarized in Figure \ref{fig:GEANT_and_pFabric_smore}. Again, FIGRET demonstrates the best performance among all the TE algorithms. By comparing Figure \ref{fig:GEANT_and_pFabric_smore} and Figure \ref{fig: GEANT pFabric}, it is evident that path selections do not fundamentally affect the effectiveness of different TE schemes. SMORE aims to enhance the robustness merely through path selection. After the paths have been determined, SMORE optimizes the TE performance based solely on the predicted traffic, without considering the potential traffic bursts. As shown by the Pred TE results in the GEANT network in Figure \ref{fig:GEANT_and_pFabric_smore}, SMORE incurs the highest tail normalized MLU among all the TE schemes. This indicates that merely enhancing TE robustness through path selection is insufficient.
\subsection{Coping with network failures}
\label{section: coping with network failures}
\begin{figure}[!ht]
    \centering
    \includegraphics[scale=0.45]{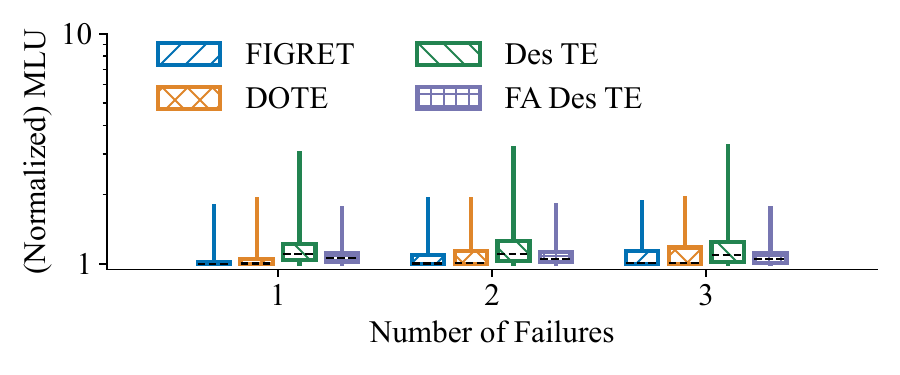}
    \caption{Coping with different numbers of random link failures on GEANT.}
    \label{fig:failure GEANT}
\end{figure}

Figure \ref{fig:failure GEANT} shows the performance of FIGRET, when different numbers of randomly selected links fail within the GEANT topology. We conduct a comparison between FIGRET, DOTE, Desensitization-based TE (Des TE), and Des TE that is also fault-aware (FA Des TE). Fault-aware refers to knowing the links that will fail in the \textit{future} and optimizing only over the paths without failures, without the need to reroute the traffic around failed paths after the TE solutions are determined. Their result is normalized against an oracle that possesses the best knowledge of both future demands and failures. Our results demonstrate that FIGRET achieves high resiliency to link failures. It outperforms both the DOTE and Des TE and is equally competitive with the FA Des TE, which has Oracle access to future failures. 
Our interpretation for this is that first, compared to DOTE, FIGRET, by incorporating robustness constraints, ensures that the traffic load on a path does not become excessive, thereby mitigating the impact of link failures on performance. The reason why FIGRET achieves comparable performance with FA Des TE is twofold: firstly, Des TE’s explicit prediction of the traffic matrix introduces errors; secondly, it lacks fine-grained robustness enhancement. These factors offset the advantage FA Des TE gains from knowing link failures.
Our results on other topologies are presented in Appendix \ref{section: additional failure results}. These additional results also exhibit similar characteristics.
\begin{table*}[h]
\centering
\begin{minipage}{.32\textwidth}
  \centering
  \scalebox{0.8}{
    \begin{tabular}{cccccc}
    \hline
    \multirow{2}{*}{Network}                    &           & \multicolumn{4}{c}{Factor $\alpha$}                                     \\
                                                &           & 0.2                       & 0.5                       & 1.0    & 2.0    \\ \hline
    \multirow{2}{*}{PoD DB}                     & average   & -0.3\%                    & 0.2\%                     & 3.2\%  & 9.8\%  \\
                                                & 90th Pct. & 2.0\%                     & 3.0\%                     & 6.6\%  & 16.7\% \\
    \multirow{2}{*}{pFabric}                    & average   & 0.2\%                     & -0.6\%                    & -1.8\% & -2.9\% \\
                                                & 90th Pct. & 0.4\%                     & 0.5\%                     & -1.0\% & 1.0\%  \\
    \multicolumn{1}{l}{\multirow{2}{*}{ToR DB}} & average   & 1.6\%                     & 3.9\%                     & 8.9\%  & 16.1\% \\
    \multicolumn{1}{l}{}                        & 90th Pct. & 0.8\%                     & 2.8\%                     & 4.5\%  & 5.3\%  \\ \hline
    \end{tabular}
    }
  \caption{Performance decline with increased traffic fluctuation. Negative values indicate no degradation.}
  \label{table: temporary changes}
\end{minipage}%
\hfill
\begin{minipage}{.32\textwidth}
    \centering
    \scalebox{0.8}{
        \begin{tabular}{ccccc}
        \hline
        \multirow{2}{*}{Network} &           & \multicolumn{3}{c}{Training data time segments} \\
                                 &           & 0\%-25\%      & 25\%-50\%      & 50\%-75\%      \\ \hline
        \multirow{2}{*}{PoD DB}  & average   & -0.6\%        & -0.4\%         & -0.4\%         \\
                                 & 90th Pct. & -0.6\%        & -0.0\%         & -0.7\%         \\
        \multirow{2}{*}{pFabric} & average   & 0.6\%         & 0.9\%          & 0.6\%          \\
                                 & 90th Pct. & 1.3\%         & 0.2\%          & 1.0\%          \\
        \multirow{2}{*}{ToR DB}  & average   & 2.5\%         & 2.0\%          & 2.5\%          \\
                                 & 90th Pct. & 3.0\%         & 1.6\%          & 1.6\%          \\ \hline
        \end{tabular}
    }
    \caption{Performance decline with natural drift in traffic. Negative values indicate no degradation.}
    \label{table: natural drift}
\end{minipage}%
\hfill
\begin{minipage}{.32\textwidth}
  \centering
  \scalebox{0.8}{
    \begin{tabular}{cccccc}
    \hline
    \multirow{2}{*}{Network} &           & \multicolumn{4}{c}{Factor}        \\
                             &           & 0.2    & 0.5    & 1.0    & 2.0    \\ \hline
    \multirow{2}{*}{PoD DB}  & average   & -0.3\% & 0.7\%  & 4.8\%  & 15.3\% \\
                             & 90th Pct. & 1.7\%  & 1.0\%  & 6.6\%  & 27.9\% \\
    \multirow{2}{*}{pFabric} & average   & 0.2\%  & -0.6\% & -1.7\% & -2.7\% \\
                             & 90th Pct. & 0.5\%  & 0.7\%  & -0.4\% & 1.6\%  \\
    \multirow{2}{*}{ToR DB}  & average   & 0.9\%  & 2.7\%  & 17.0\% & 38.6\% \\
                             & 90th Pct. & 0.0\%  & 2.6\%  & 23.1\% & 32.4\% \\ \hline
    \end{tabular}
 }
 \caption{Performance decline under worst-case conditions. Negative values indicate no degradation.}
\label{table: worst case performance}
\end{minipage}
\end{table*}
\subsection{Robustness to demand changes}
\label{section:robustness to demand changes}
\textbf{Temporal changes in traffic.} We test FIGRET for traffic variability. Our method for increasing variability is as follows: For each source-destination (SD) pair in the real traffic, we generate a traffic fluctuation using the Gaussian distribution \(N(\mu, \sigma^2)\) multiplied by a factor \(\alpha\), where \(\mu = 0\) and \(\sigma\) is the standard deviation of the traffic demand $d_{st}$. Here, \(\alpha\) is chosen from the set \{0.2,0.5,1.0,2.0\}, indicating the amplitude of the fluctuation. We observe how much the performance of FIGRET decreased after imposing sudden traffic changes, compared to its performance without such changes. The results are summarized in Table \ref{table: temporary changes}. It can be observed that when the factor \(\alpha\) is relatively small, there is no significant decrease in FIGRET's performance. At the same time, when the factor \(\alpha\) is 2, doubling the noise of \(N(0, \sigma^2_{D_{sd},[1-T]})\), the performance does not decrease by more than \(20\%\).\\
\noindent\textbf{Natural drift in traffic.} We test the impact of natural traffic shifts on FIGRET. In the aforementioned experiment, we sorted the data chronologically, using the first 75\% for training and the latter 25\% for testing. However, in this section, we conduct training separately with 0\%-25\%, 25\%-50\%, and 50\%-75\% of the data, followed by testing on the last 25\%. This approach allows us to observe the performance decline compared to when 75\% of the data is used for training. We focus on two aspects: firstly, the impact of training with less data on FIGRET's performance; secondly, the effect of the update frequency on performance. The results are summarized in Table \ref{table: natural drift}.
It is observed that FIGRET's performance remains largely unaffected even a long time after training completion (exceeding two times the total duration of the training data).
To fully understand how traffic data evolves across time, we also visually analyzed the data distribution across various times, presented in Appendix \ref{section:visualization of traffic demands}. The analysis indicates that neither PoD-level nor ToR-level data exhibit drastic changes in their traffic patterns over time. However, the shift effect at the ToR level is somewhat more pronounced than at the PoD level. This corresponds with our results in Table \ref{table: natural drift}. Through this analysis, we are inspired that the FIGRET training does not necessarily need to be especially frequent.\\
\noindent\textbf{Worst-case performance.}
We evaluate the performance of FIGRET under worst-case conditions. Considering that FIGRET develops strategies based on the characteristics of historical data, we intentionally reverse the order of the magnitude of temporal traffic fluctuations among SD pairs, applying larger traffic fluctuations to those pairs with historically lower variances. The results are summarized in Table \ref{table: worst case performance}. Compared to the results in Table \ref{table: temporary changes}, although there is indeed a greater decline in FIGRET’s performance, with PoD DB and ToR DB experiencing a 30\%-40\% reduction when $\alpha=2$, the performance did not completely fail. Additionally, we employ the Spearman rank correlation coefficient to assess the consistency of variance ranking changes between the test and train sets across different SD pairs within our dataset. The analysis reveals a high Spearman rank correlation coefficient, with PoD DB at 0.92 and ToR DB at 0.98. These results suggest that the actual occurrence of the worst-case scenario we are considering is rare. Furthermore, for the pFabric system, since the sources and destinations of traffic are selected through a uniform random process, the differences in variance among SD pairs are not pronounced. Therefore, the performance variations are not significant.
\subsection{Interpreting FIGRET's effectiveness}
\label{section: interpretation of termol}
Beyond the impressive end-to-end results achieved by FIGRET, we aim to provide a high-level explanation that is easily understandable to humans. In this way, we aim to make network operators more inclined to implement FIGRET in practice.
\begin{figure}[!h]
    \subfigure[Hedge-based TE, PoD-level, Meta DB cluster.]{
    \label{fig:jupiter facebook_pod_a}
    \includegraphics[scale=0.18]{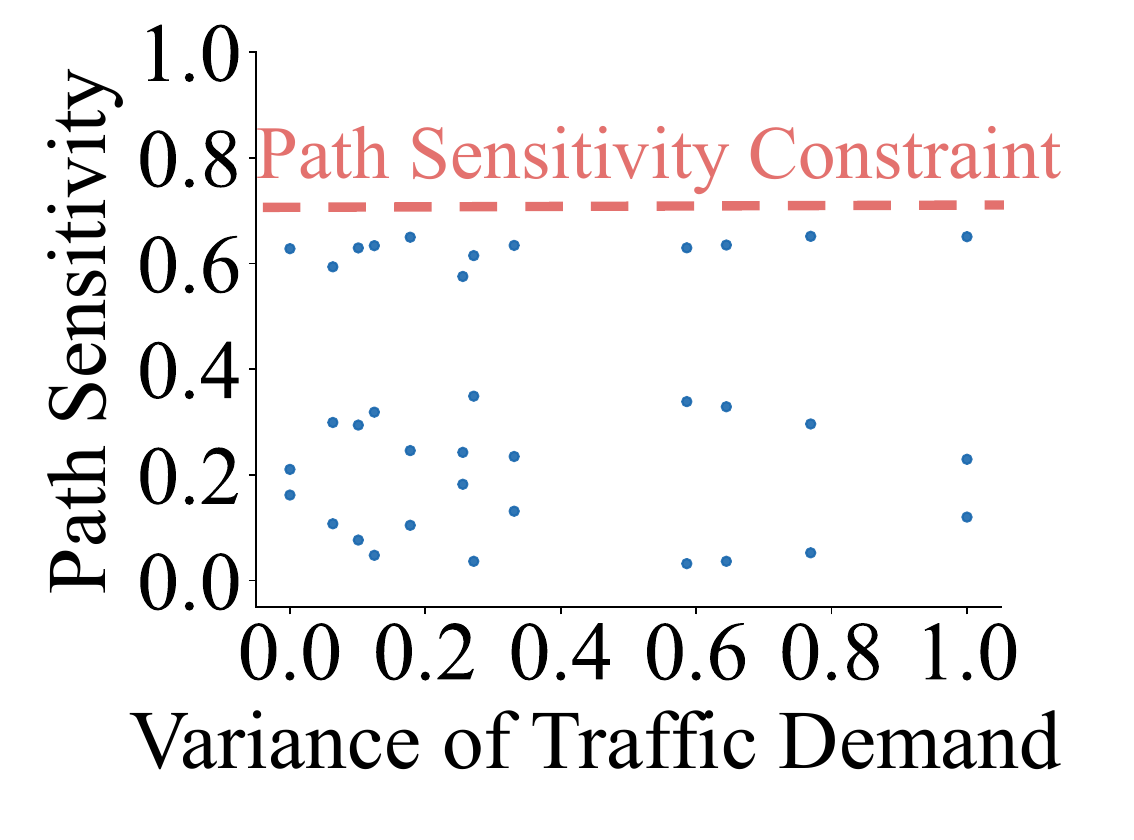}
    }
    \subfigure[FIGRET, PoD-level, Meta DB cluster.]{
    \label{fig:figret facebook_pod_a}
    \includegraphics[scale=0.18]{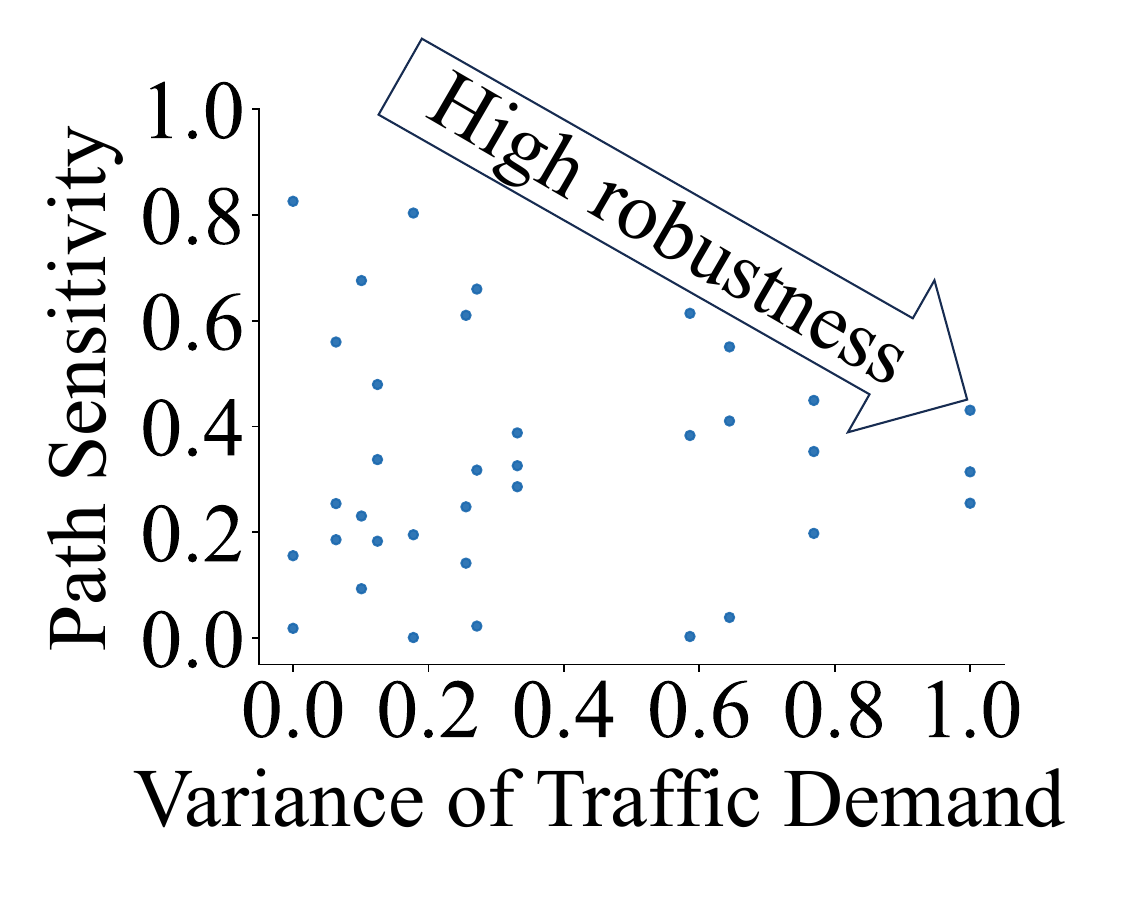}
    }
    \subfigure[Hedge-based TE, ToR-level, Meta DB cluster.]{
    \label{fig:jupiter facebook_tor_a}
    \includegraphics[scale=0.18]{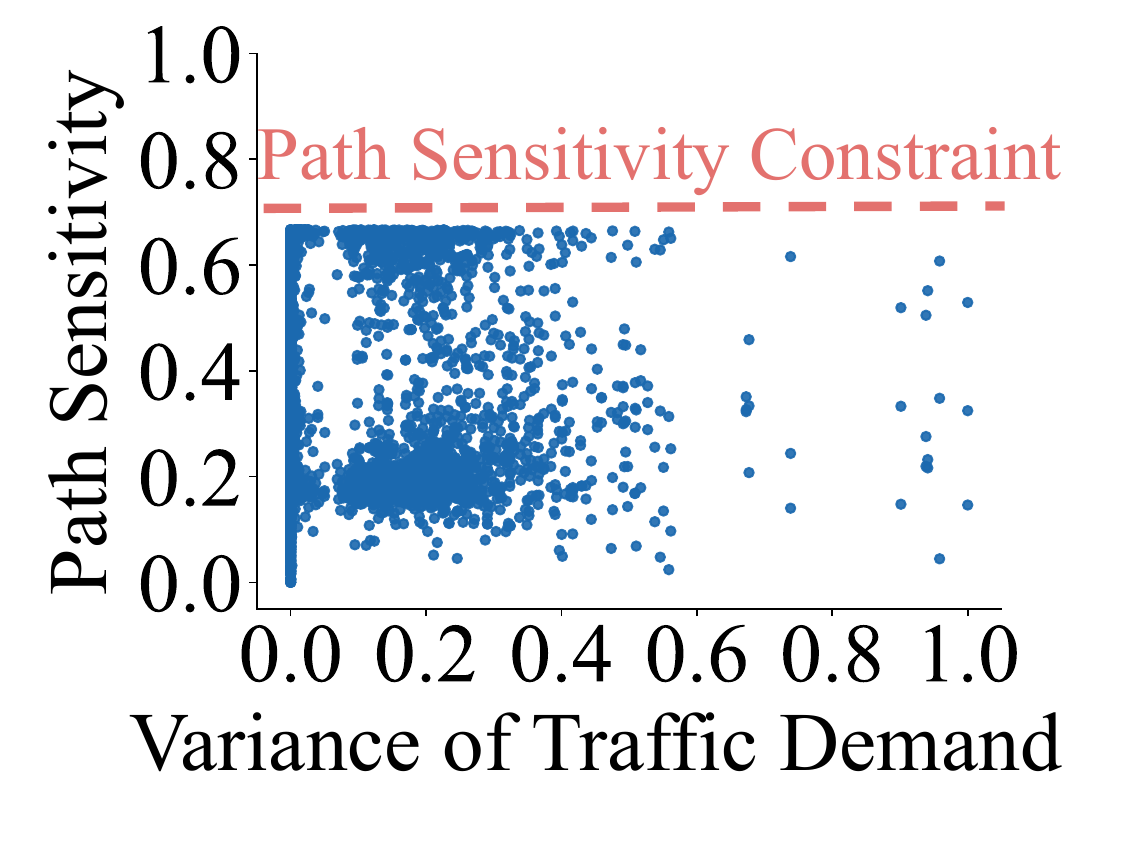}
    }
    \subfigure[FIGRET, ToR-level, Meta DB cluster.]{
    \label{fig:figret facebook_tor_a}
    \includegraphics[scale=0.18]{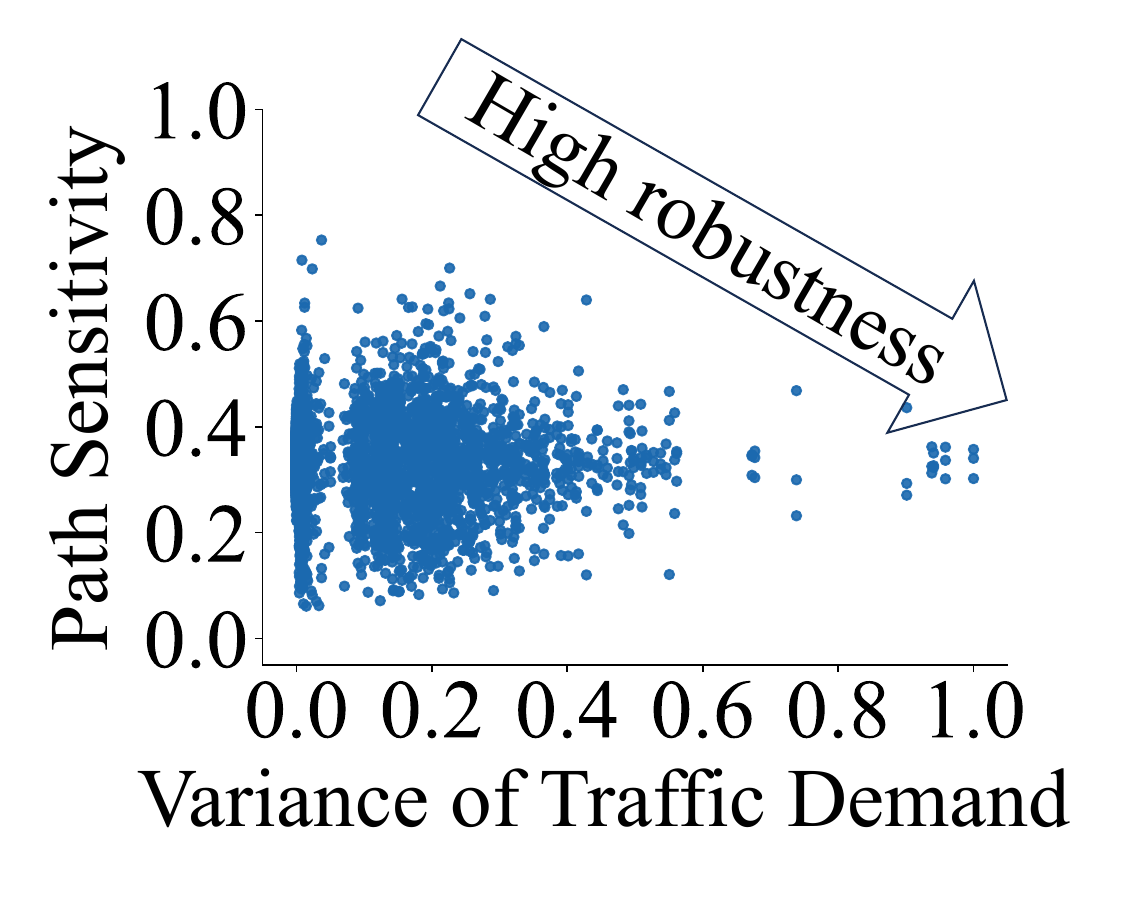}
    }
    \caption{Compare the distribution of path sensitivities in Hedge-based TE with FIGRET. Each point in the graph represents a path, with the x-axis indicating the variance of the traffic served by that path. The y-axis represents the average sensitivity of the path across all tested traffic demands.}
    \label{fig: interpret termol}
\end{figure}

\noindent\textbf{Sensitivity analysis in FIGRET}. We numerically examine how FIGRET achieves fine-grained robustness. We compare the network configurations of Google's hedge-based TE and FIGRET, examining the relationship between path sensitivity and traffic characteristics. The results are presented in Figure \ref{fig: interpret termol}. When creating Figure \ref{fig: interpret termol}, we normalize the variance of the traffic on the x-axis. For the y-axis, when calculating path sensitivity, we normalize the capacity of the edges (considering the edge with the smallest capacity in the topology as 1). Furthermore, for the ToR level topology in Figure \ref{fig:jupiter facebook_tor_a} and \ref{fig:figret facebook_tor_a}, with 155 nodes representing $155\times154$ SD pairs and three paths per SD pair, there are a total of 1551543 paths. Due to the excessive number of points to plot, we uniformly select $10\%$ of the SD pairs for better visualization.
From these results, we have the following observations: 1) For hedge-based TE, the path sensitivity of all paths, whether traffic is bursty or not, is constrained within a constant upper bound. 2) For FIGRET, path sensitivity settings depend on traffic characteristics: it's dispersed for non-bursty and concentrated for bursty traffic. This enables fine-grained robustness analysis, achieving a better balance between normal-case performance and robustness to burst traffic.
\section{discussion}
\textbf{Is deep learning essential for implementing the concept of fine-grained robustness enhancement?} In the design of FIGRET, we harness deep learning to devise a robustness enhancement scheme based on complex traffic characteristics and network topologies. However, considering that commercial network operators may be reluctant to overhaul their existing in-production TE systems completely, we discuss in this section whether slight modifications to the existing TE schemes, incorporating the principles of robustness enhancement, could result in better performance compared to the original TE schemes. To investigate this, we make slight modifications to the TE system currently in use by Google's Jupiter data centers. We discover that by designing simple heuristic-based $\mathcal{F}$ path sensitivity constraint functions that align with fine-grained robustness concepts, the performance of TE algorithms can be improved. We use linear functions to impose stricter constraints as traffic variability increases. We also use piecewise functions to divide traffic into more stable and bursty segments, applying looser constraints to stable segments and stricter constraints to bursty segments. We validate that even such simple functions can improve the original TE’s performance. For example, when using linear functions, if we maintain the constraints for high bursty traffic but relax the constraints for low bursty traffic, we can achieve a 5\% reduction in the average normal-case MLU while maintaining robustness (see Appendix \ref{section: heuristic selection} for more details).

Although it is feasible to directly apply fine-grained robustness enhancements to traditional TE systems currently in operation, this approach encounters two significant issues: firstly, the use of LP methods for solving is notably slow. Secondly, manually configured functions often fall short of achieving optimal performance due to their inability to fully adapt to complex networks and dynamic traffic demands. Given these considerations, this paper employs a method based on deep learning. \\
\textbf{When should FIGRET be retrained?} In this paper, FIGRET adopts a relatively simple periodic retraining approach, and as demonstrated in \S \ref{section:robustness to demand changes}, retraining does not necessarily have to be particularly frequent. Thanks to FIGRET's rapid training times, this periodic training approach can meet usage requirements. Alternative methods, such as retraining after detecting significant changes in network traffic patterns or a certain degree of performance degradation, might more accurately determine the timing for retraining. However, we leave these considerations for future work.\\
\textbf{Is it feasible to deploy deep learning-based TE methods in practice?} In general, there are two main obstacles that may prevent the practical deployment of deep learning-based methods: 1) the learning-based methods may sometimes produce incorrect or infeasible solutions, and 2) the performance results are not easily interpretable, raising concerns about the solution reliability in different cases. Nevertheless, we argue that the above issues are less of a concern for FIGRET. First, the constraints of TE, which require that the sum of proportions served by all paths for a given source-destination pair equals one, can be easily enforced by normalizing the outputs of the neural network, thus avoiding infeasible solutions. Second, FIGRET's superior performance is explainable, as demonstrated in \S\ref{section: interpretation of termol}.\\
\textbf{Can the concept of fine-grained robustness be extended to other objectives?} Firstly, fine-grained robustness in traffic not only applies to minimizing MLU but can also be extended to other objectives such as minimizing latency. For instance, selecting the shortest path for non-burst traffic can achieve minimal latency; however, for potential burst traffic, employing multipath transmission may be necessary to avoid congestion on the shortest paths. Secondly, the concept of fine-grained robustness is also applicable to handling link failures. Existing strategies, such as FFC \cite{liu2014traffic}, often optimize for all possible failure scenarios, which can lead to low utilization and over-provisioning. By adopting a fine-grained approach that considers the probability of failure scenarios, resource utilization, and fault management can be balanced more effectively.
\section{related work}
\textbf{Robustness-enhanced-based TE.}
A major focus of research on TE is to improve the algorithm's robustness in handling sudden traffic bursts. Oblivious TE \cite{applegate2003making} aims to enhance robustness by consistently considering the most congested scenario, optimizing for the worst-case MLU among all possible DMs. COPE \cite{wang2006cope} optimizes MLU across a set of DMs predicted by historical DMs while holding a worst-case MLU guarantee. Desensitization-based TE \cite{poutievski2022jupiter,teh2020couder}, by introducing the concept of path sensitivity which characterizes the impact of traffic bursts on the path, aims to enhance robustness by optimizing the MLU while ensuring low path sensitivity. Robustness-enhanced TE schemes typically exhibit coarse granularity by treating each source-destination pair in the traffic matrix equally. In contrast, FIGRET stands apart by employing fine-grained robustness enhancement, resulting in a better balance between robustness and average-case performance.\\
\textbf{Threshold-based VLB Routing}: TROD \cite{cao2021trod, cao2023threshold} introduces a threshold-based routing strategy in optical data centers to handle unexpected traffic bursts. The basic idea is that when traffic demand is below a predetermined threshold, the traffic is routed via shortest paths. When traffic demand exceeds the threshold, the burst traffic is load-balanced across all non-shortest paths. TROD maintains normal-case performance by routing most traffic via the shortest paths, while it enhances robustness by dispersing bursts across multiple paths. Hence, TROD strikes a good balance between normal-case and worst-case performance. Unfortunately, TROD's routing algorithm requires special hardware support. P4 switches can support TROD's routing algorithm, but they are not widely deployed in data centers. In contrast, FIGRET does not require specialized hardware and only needs switches that support WCMP.\\
\textbf{Machine learning-based TE.} 
Machine learning-based TE can be broadly categorized into two classes. The first class is Demand Prediction-based TE [1, 31, 35, 36]. These methods use predictive models to estimate the next traffic matrix (TM) and then perform optimization based on the estimated TM. The second class replaces explicit demand prediction with end-to-end optimization, directly mapping a recent historical window of TMs to TE configurations \cite{perry2023dote,valadarsky2017learning}. Despite harnessing the powerful capabilities of deep learning to learn relationships between traffic flows, both of these classes overlook the consideration of unexpected traffic bursts. \\
\textbf{Network planning.} 
The objective of network planning is to configure network capacity to ensure good performance for all possible traffic patterns. Consequently, Meta models the traffic based on the Hose model \cite{eason2023hose,ahuja2021capacity} and optimizes for the worst-case performance. This approach is not directly applicable to TE because it adjusts network capacity rather than routing weights on paths. Furthermore, even if one may repurpose a network planning approach to solve TE problems, the resulting solution lacks the capability to balance normal-case performance and burst-case performance.
\section{conclusion}
In this work, we introduce FIGRET, a new design point for TE that enhances robustness in a fine-grained manner based on traffic characteristics. FIGRET employs path sensitivity to manage traffic bursts and utilizes a deep learning strategy with a well-designed loss function to produce TE solutions that consider robustness at a fine-grained level. Our experimental results demonstrate that FIGRET achieves an effective balance between normal-case performance and robustness, resulting in high-quality TE solutions.

\section*{ACKNOWLEDGMENTS}
We sincerely thank our shepherd Manya Ghobadi and the anonymous reviewers for their constructive feedback. This work was supported by the NSF China (No. 62272292, No. 62132009 and No. 61960206002).

\newpage
\bibliographystyle{ACM-Reference-Format}
\bibliography{reference}


\appendix
\clearpage
\section*{Appendix}
Appendices are supporting material that has not been peer-reviewed.
\section{Notation table}
In this section, we tabulate the notations in Table \ref{table: notations used in this paper}.
\begin{table}[h]
\begin{tabular}{|l|l|}
\hline
\textbf{Notation}               & \textbf{Description}                                                                                                                                                     \\ \hline
$G(V,E,c)$                      & \begin{tabular}[c]{@{}l@{}}Network topology, $V$ is vertex set, $E$ is edge \\ set, and c assigns capacities to edges\end{tabular}                                       \\ \hline
$D$                             & \begin{tabular}[c]{@{}l@{}}Demand matrix, where $D_{ij}$ denotes the\\ traffic from $i$ to $j$\end{tabular}                                                              \\ \hline
$P$                             & \begin{tabular}[c]{@{}l@{}}Network paths, where $P_{sd}$ denotes the set \\ of network paths through which source \\ $s$ communicates with  destination $d$\end{tabular} \\ \hline
$C_p$                           & The capacity of path $p$                                                                                                                                                 \\ \hline
$r_p$                           & \begin{tabular}[c]{@{}l@{}}The split ratio of the traffic demand from \\ $s$ to $d$ forwarded along path $p$\end{tabular}                                                \\ \hline
$\mathcal{R}$                   & TE configuration                                                                                                                                                         \\ \hline
$M(\mathcal{R},D)$              & MLU in the network given $D$ and $\mathcal{R}$                                                                                                                           \\ \hline
$\Delta$                        & The set representing mismatch                                                                                                                                            \\ \hline
$\mathscr{S}_p$                 & Path sensitivity for path $p$                                                                                                                                            \\ \hline
$\mathscr{S}_{sd}^{\text{max}}$ & \begin{tabular}[c]{@{}l@{}}Maximum path sensitivity among all paths\\ in $P_{st}$\end{tabular}                                                                           \\ \hline
$\mathcal{F}$                   & Function for path sensitivity constraints                                                                                                                                \\ \hline
$\pi_\theta$                    & \begin{tabular}[c]{@{}l@{}}The mapping function from DNN inputs to \\ outputs, and $\theta$ denotes the DNN's weight\end{tabular}                                 \\ \hline
$H$                             & \begin{tabular}[c]{@{}l@{}}Window length, representing the number of \\ historical traffic demands in the DNN input\end{tabular}                                      \\ \hline
$\mathcal{L}(\mathcal{R}_t,D_t)$& Loss function                            \\ \hline
$\sigma^2_{D_{sd},[1-T]}$       & \begin{tabular}[c]{@{}l@{}}The variance of traffic demands $D_{sd}$ within \\ the time range from $1$ to $T$\end{tabular}                                                 \\ \hline
\end{tabular}
\caption{Notations used in this paper}
\label{table: notations used in this paper}
\end{table}
\section{TE optimization formulation}
\label{section: te opt formulation}
In this section, we formulate the TE optimization problem. Given a network $G(V,E,c)$ and traffic demand $D$, to find an optimal TE configuration which specifies a split ratio $r_p$ for each path $p \in P_{st}$, where $r_p$ represents the fraction of the traffic demand from $s$ to $t$ forwarder along path $p$, the optimization formulation shown in Equation \ref{equation: te opt formulation} can be established.
\begin{equation}
\boxed{
\label{equation: te opt formulation}
\begin{aligned}
&\mathop{\text{minimize}}\limits_{\mathcal{R}}\qquad\max_{e\in E} \frac{f_e}{c(e)}\\
&\text{subject to}\qquad\sum_{p \in P_{st}} r_p &&= 1,\forall s,t\in V\\
&\text{          }\qquad\sum_{s,t \in V, p\in P_{st},e\in p}D_{st} \times r_p &&= f_e ,\forall e\in E
\end{aligned}
}
\end{equation}
Our work, however, focuses on realistic scenarios where TE schemes must select configurations on the realistic scenario in which the traffic demand is not known beforehand.
\section{Heuristic selection of $\mathcal{F}$}
\label{section: heuristic selection}
In this section, we integrate the Desensitization-based TE scheme in Google's Jupiter Evolving \cite{poutievski2022jupiter}, replacing its fixed path sensitivity constraints with a heuristically selected function $\mathcal{F}:(s,t) \rightarrow \mathbb{R}^+$. As discussed in \S \ref{section: TERMOL design}, for stable traffic, we apply lenient sensitivity constraints, whereas, for highly bursty traffic, we impose stringent constraints. We use traffic variance as an indicator of traffic burstiness. Therefore, the path sensitivity constraints should become increasingly stringent as the path variance increases, meaning the maximum allowable path sensitivity should decrease correspondingly. We experiment with two heuristic functions: 1) Linear function, and 2) Piecewise function.\\
\subsection{Linear function} 
\label{section: linear function}
\begin{figure}[!h]
    \centering
    \includegraphics[scale=0.3]{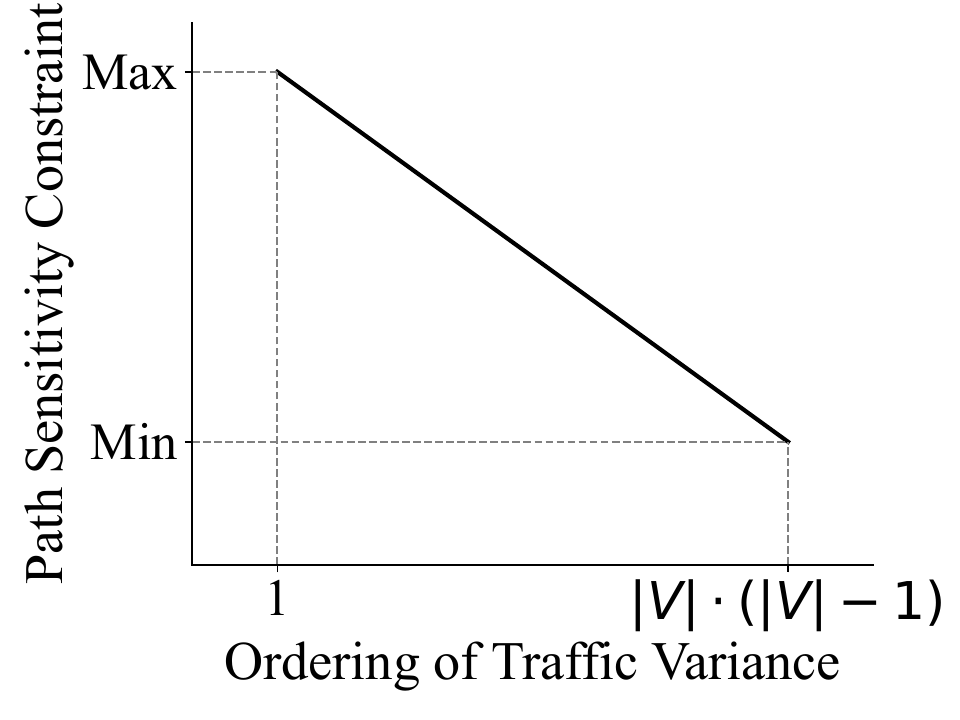}
    \caption{Illustrating the function $\mathcal{F}$ selected by Linear function.}
    \label{fig:function f liner}
\end{figure}

The function we have selected is illustrated in Figure \ref{fig:function f liner}. Using a linear function, we arrange the traffic variance for all SD pairs in ascending order and assign varying path sensitivity constraints based on different orders. The \textit{Max} and \textit{Min} values in the Figure can be arbitrarily chosen, but ensuring that a solution is feasible when selecting Min is crucial. For instance, if there are $n$ paths serving this SD pair, each with capacity of $1$, it is required that $r_p \leq \text{Min}$ and $\sum_{i=1}^n r_i =1$. In this scenario, Min should not be less than $\frac{1}{n}$; otherwise, the problem becomes unsolvable.

We conduct tests using varying \textit{Min} and \textit{Max} parameters, which are listed in Table \ref{table: linear function}. Strategy 1 refers to the implementation of stricter path sensitivity restrictions. In contrast, Strategy 2 involves relaxing the path sensitivity restrictions for more stable flows. Meanwhile, Strategy 3 suggests a combination approach: it proposes enforcing stricter path sensitivity restrictions on sudden or bursty flows while simultaneously relaxing these restrictions for stable flows. For ease of numerical representation, we have normalized the capacities of all paths in the graph, setting the smallest capacity to 1, with other capacities calculated proportionally.
\begin{table}[!h]
\resizebox{0.45\textwidth}{!}{
\begin{tabular}{cccccc}
\hline
       & \multicolumn{2}{c}{Strategy 1} & Original & Strategy 2    &  Both   \\ \hline
Number & 1         & 2        & 3        & 4   & 5   \\
Min    & 1/3$\downarrow$       & 1/3$\downarrow$      & 2/3      & 2/3 & 1/3$\downarrow$ \\
Max    & 1/2$\downarrow$       & 2/3      & 2/3      & 5/6$\uparrow$ & 5/6$\uparrow$ \\ \hline
\end{tabular}
}
\vspace{3mm}
\caption{The parameters selected for the Linear function. `Original' refers to the fixed path sensitivity constraint of the Desensitization-based TE.}
\label{table: linear function}
\end{table}

\begin{figure}
    \centering
    \includegraphics[scale=0.3]{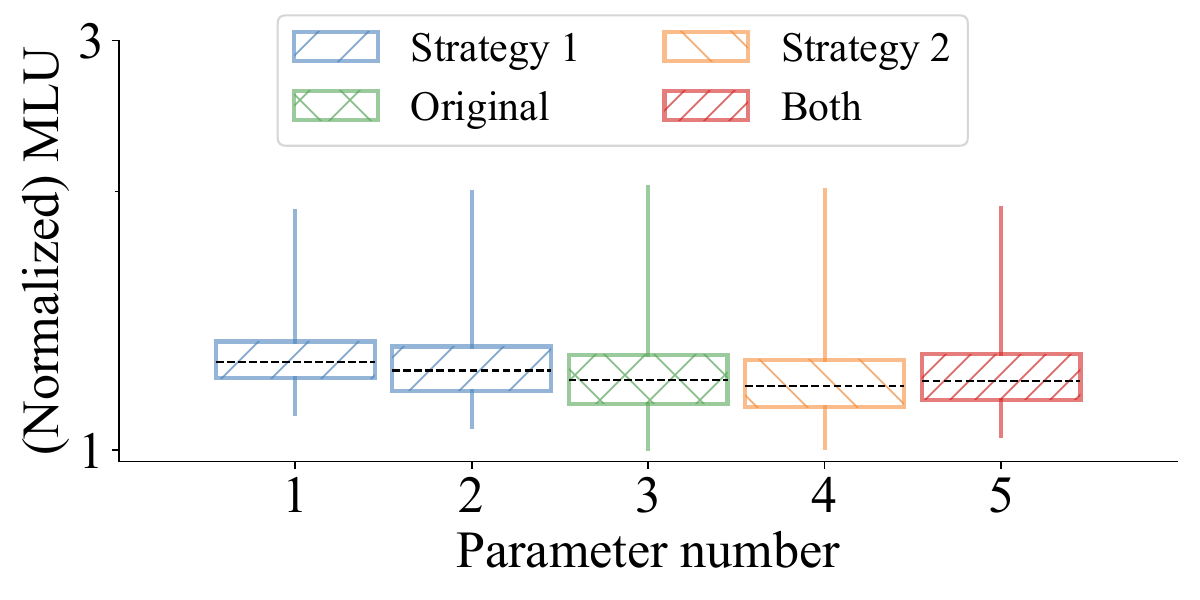}
    \caption{Comparing the TE quality under different parameter settings in Linear function. The parameter number on the x-axis corresponds to the numbering in Table \ref{table: linear function}}
    \label{fig:linear te quality}
\end{figure}

We apply the parameters listed in Table \ref{table: linear function} and conduct tests on the PoD-level Meta DB. The results are summarized in Figure \ref{fig:linear te quality}. As shown in Figure \ref{fig:linear te quality}, the application of Strategy 1, which involves imposing stricter path sensitivity constraints (as in the case of the group \{1,2,3\}), leads to an enhanced capability to handle bursty traffic. On the other hand, with the implementation of Strategy 2, an improvement in average performance is observed, as demonstrated by the comparison of groups \{3,4\}. Comparing parameters 5 with parameters 1, we observe that by maintaining the sensitivity constraint for high bursty traffic while relaxing the sensitivity constraint for non-bursty traffic, parameters 5 can reduce the normal-case performance (below the 75th percentile) MLU by 5\% at the expense of increasing the peak MLU by 0.7\%.
\subsection{Piecewise function}
\label{section: piecewise function}
\begin{figure}[!h]
    \centering
    \includegraphics[scale=0.3]{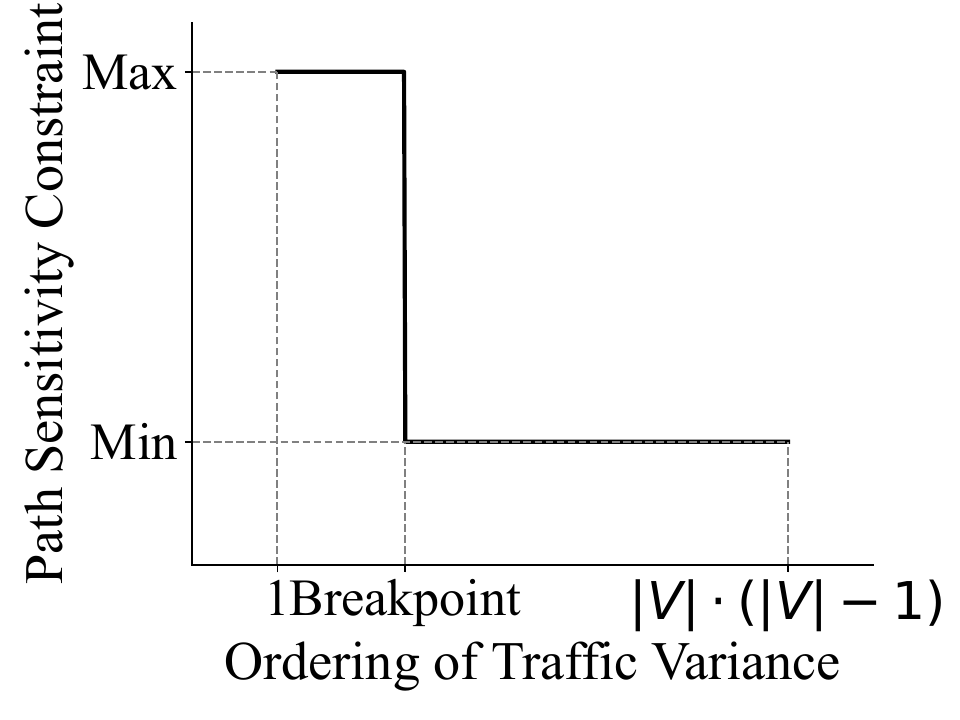}
    \caption{Illustrating the function $\mathcal{F}$ selected by Piecewise function.}
    \label{fig:function f step}
\end{figure}

The function we have selected for our analysis is depicted in Figure \ref{fig:function f step}. It utilizes a piecewise approach to define constraints on path sensitivity, following the same x-axis and y-axis interpretations as shown in Figure \ref{fig:function f liner}. We specifically chose a piecewise function for this purpose. We introduce a breakpoint within the function to distinguish between stable and bursty traffic conditions. For traffic variances that fall below this breakpoint, indicative of stable flow, the path sensitivity constraints are comparatively relaxed. In contrast, for traffic variances that exceed this breakpoint, representing bursty flow, the constraints are tighter and more rigorous.

We employ various parameters to demonstrate the impact of different parameter settings on the results. In this function, the parameters available for selection include \textit{Min}, \textit{Max}, and \textit{breakpoint}. Our parameter settings are presented in Table \ref{table: piecewise parameter}. For ease of numerical representation, we have normalized the capacities of all paths in the graph, setting the smallest capacity to 1, with other capacities calculated proportionally. Additionally, we represent the value of the breakpoint in terms of its proportion rather than as an absolute number. For example, $\text{breakpoint=0.8}$ signifies that the first $80\%$ of the flow is relatively stable, while the latter $20\%$ comprises bursty traffic.
\begin{table}[!ht]
\resizebox{0.45\textwidth}{!}{
\begin{tabular}{cccccccc}
\hline
           & \multicolumn{3}{c}{Strict Constraint} & Original         & \multicolumn{3}{c}{Relaxed Constraint}                          \\ \hline
Number     & 1                 & 2     & 3    & 4                & 5   & \multicolumn{1}{c}{6}    & \multicolumn{1}{c}{7}   \\
Min        & 1/2$\downarrow$   & 1/2$\downarrow$   & 1/2$\downarrow$  & 2/3              & 2/3 & 2/3                      & 2/3                     \\
Max        & 2/3               & 2/3   & 2/3  & 2/3              & 5/6$\uparrow$ & 5/6$\uparrow$                      & 5/6$\uparrow$                     \\
Breakpoint & 0.5               & 0.65  & 0.8  & \textbackslash{} & 0.5 & \multicolumn{1}{c}{0.65} & \multicolumn{1}{c}{0.8} \\ \hline
\end{tabular}
}
\vspace{3mm}
\caption{The parameters selected for the Piecewise function. `Original' refers to the fixed path sensitivity constraint of the Desensitization-based TE.}
\label{table: piecewise parameter}
\end{table}

\begin{figure}[!ht]
    \centering
    \includegraphics[scale=0.3]{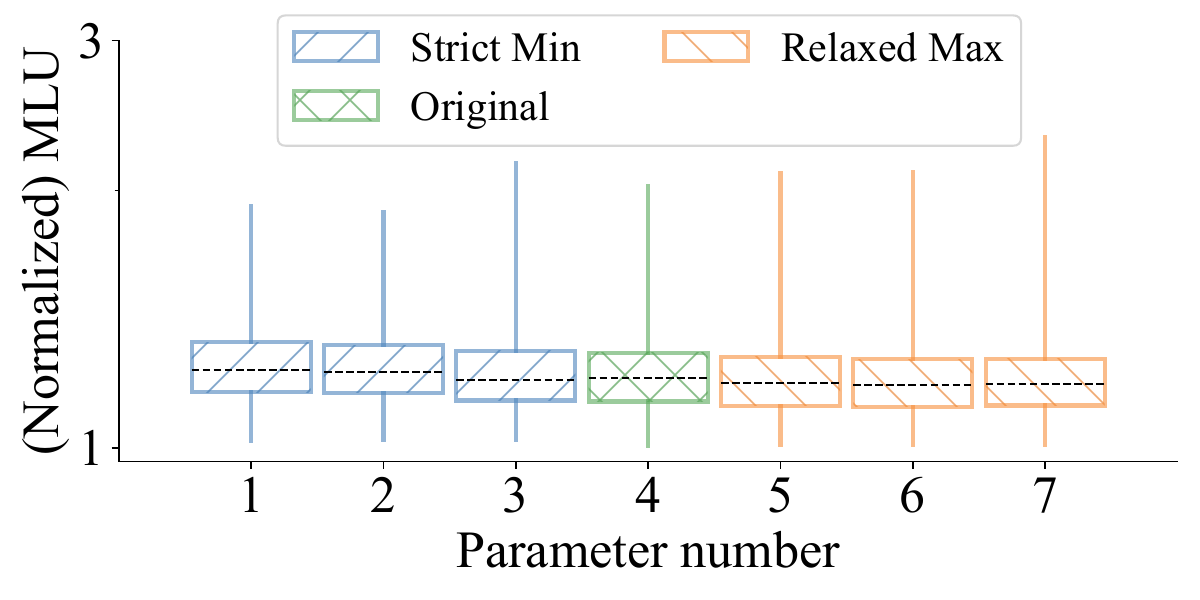}
    \caption{Comparing the TE quality under different parameter settings in Piecewise function. The parameter number on the x-axis corresponds to the numbering in Table \ref{table: piecewise parameter}.}
    \label{fig:step te quality}
\end{figure}

In our study, we conducted a series of tests on the parameters listed in Table \ref{table: piecewise parameter} on PoD-level Meta DB, with the results summarized in Figure \ref{fig:step te quality}. As illustrated in the figure, with fixed values of Min and Max, a larger Breakpoint setting enhances the average performance of TE methods. This trend is evident in groups \{1,2,3\} and \{5,6,7\}, where an increase in Breakpoint corresponds to a lower average performance boxplot. Conversely, with a constant Breakpoint, maintaining Max while reducing Min enhances the TE methods' ability to manage bursty traffic, as shown in the comparison of group \{1,4\}. Similarly, fixing Min while increasing Max leads to better average performance of TE methods, as observed in the comparison between groups \{4,5\}.

These findings align with intuitive expectations: a higher proportion of stable traffic or more lenient path sensitivity constraints for stable traffic result in improved average algorithm performance. Conversely, a lower proportion of stable traffic with stricter constraints for bursty traffic enhances the algorithm's ability to handle sudden bursts. 
\subsection{Summary} 
This section demonstrates that by implementing fine-grained robustness enhancement strategies into Google's TE using simple heuristic functions, we can achieve varying effects through parameter adjustments, thereby enhancing the original TE algorithm's capabilities. Our application of deep learning in FIGRET aims to identify the function form and parameters for fine-grained robustness enhancement as effectively as possible, while also speeding up the TE solution process.
\section{FIGRET Implementation Details}
\label{section: dnn architecture details}
While we have chosen FCN among FCN, CNN, and GNN for our application, this does not imply that TE problems must exclusively use FCNs. More sophisticated and advanced network architectures can be further explored.
\subsection{No need for GNN}
\label{section: no need for GNN}
The mapping between TE configuration and Maximum Link Utilization (MLU) can be represented using Function \ref{algorithm: mapping MLU}. As seen from Function \ref{algorithm: mapping MLU}, this mapping can be established through simple matrix operations.
\begin{algorithm}[!ht]
\caption{Mapping traffic configurations to MLU}
\label{algorithm: mapping MLU}
$G=(V,E,c)$ \tcp{\textrm{\textit{Graph that models the network topology}}}
$\Omega =\{(i,j)|i\in V,j\in V, i\neq j\}$ \tcp{\textrm{\textit{the sef of all source-destination (SD) pairs}}}
$\Phi=\bigcup\limits_{{s,t}\in \Omega}P_{st}$ \tcp{\textrm{\textit{the set of all paths}}}
$SDtoPath^{|\Omega| \times |\Phi|}$ \tcp{\textrm{\textit{Signifies whether path j serves the SD pair i. $SDtoPath_{i,j} = 1$ if path j serves SD pair i, 0 otherwise}}}
$PathtoEdge^{|\Phi| \times |E|}$ \tcp{\textrm{\textit{Signifies whether path $i$ contains edge $j$. $PathtoEdge_{i,j}=1$ if edge $j$ $\in$ path i, 0 otherwise}}}
$\mathcal{R}^{|\Phi| \times 1}$ \tcp{\textrm{\textit{Network configuration yields the split ratios for all paths}}}
$C^{|E|\times1}$ \tcp{\textrm{\textit{vector representing link capacities}}}
$DM_{|\Omega|\times 1}$ \tcp{\textrm{\textit{Flatten traffic demand matrix}}}
\tcc{\textrm{\textit{Compute MLU form $\mathcal{R}$. $\times$ for matrix multiplication; $\odot$ for element-wise (Hadamard) multiplication}}}
$FlowOnPath^{|\Phi|\times 1} = SDtoPath^T \times DM \odot \mathcal{R}$ \tcp{\textrm{\textit{$(|\Phi|,|\Omega|)\times(|\Omega|,1)\odot (|\Phi|,1)$, calculate the flow on each path}}}
$FlowOnEdge^{|E|\times1}$ = $PathtoEdge ^T \times FlowOnPath$
\tcp{\textrm{\textit{$(|E|,|\Phi|)\times(|\Phi|,1)$, calculate the flow on each edge}}}
$MLU = Max(FlowOnEdge/C)$
\end{algorithm}
\subsection{The inappropriateness of CNN}
\begin{figure}[!h]
    \subfigure[Network topology]{
    \label{fig: CNN topo}
    \includegraphics[scale=0.36]{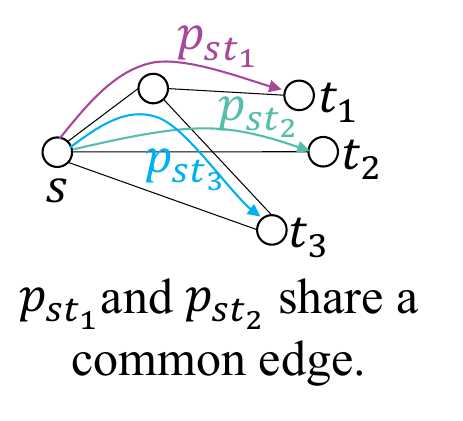}
    }
    \subfigure[Diagram illustrating convolutional operation]{
    \label{fig:cnn drawback}
    \includegraphics[scale=0.37]{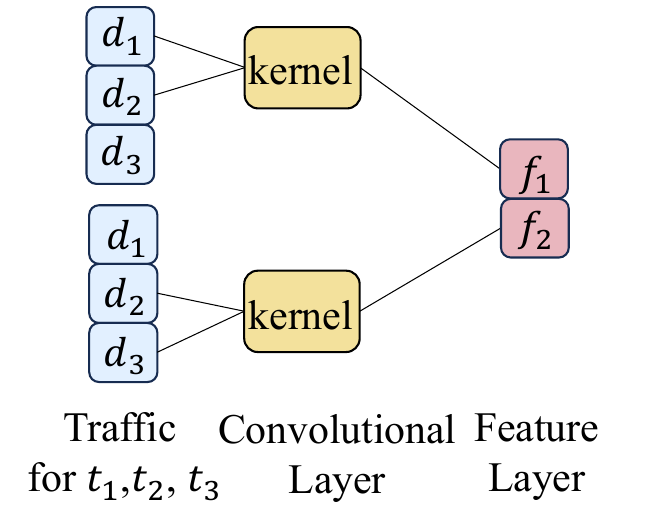}
    }
    \caption{Diagram illustrating the convolution operation in TE. Traffic demands for paths that share a common edge need to be considered jointly due to their potential impact on each other. In contrast, for paths without a shared edge, jointly considering their traffic demands is ineffective.}
    \label{fig: cnn}
\end{figure}

In Figure \ref{fig: cnn}, there is a source $s$ and three destinations $t_1, t_2, t_3$, with a convolution kernel size of 2. This kernel aims to extract \textit{local} information of length 2 from the traffic demands. However, extracting local information between ${d_1, d_2}$ and ${d_2, d_3}$ proves ineffective because $p_{st_1}$ and $p_{st_2}$, as well as $p_{st_2}$ and $p_{st_3}$, do not share a common edge. Conversely, $p_{st_1}$ and $p_{st_3}$ do share a common edge, and their traffic demands should be considered jointly. This limitation cannot be addressed by such convolution layers that are designed to extract local information. Unlike image data, which CNNs excel at handling due to the localized relevance of information, TE problems do not exhibit a distinct need for extracting local information because traffic demands within a convolutional locality may not necessarily pass through a common edge, making the use of convolution for local information extraction minimally effective. Therefore, in our model design, we chose not to incorporate convolutional layers. Instead, we opted for directly utilizing fully connected layers.
\subsection{Drawbacks of Using RL}
Reinforcement learning is extensively applied in Traffic Engineering (TE), as exemplified by TEAL \cite{xu2023teal}. This method learns strategies for achieving objectives in complex environments through trial and error. However, it often comes with high computational complexity and sensitivity to parameter settings. In TE, the relationship between policies—TE configurations and the objective of minimizing link utilization (MLU) is quite explicit and can be directly expressed. In such relatively simple scenarios, reinforcement learning may not be the most optimal choice. Under these circumstances, it's unnecessary to undergo the complex exploration and trial-and-error adjustments typical of reinforcement learning. Instead, direct application of gradient descent methods could be more effective.

In Table \ref{table: computing and precomputation}, we present a comparison of the training times for FIGRET and TEAL, which involve GNN and reinforcement learning, as well as the precomputation time for Oblivious and COPE. It is evident that as the network topology scales up, FIGRET's precomputation time offers a greater advantage.
\subsection{Implementing FIGRET}
\label{section: implementing FIGRET}
\textbf{DNN architecture.} Similar to DOTE, except for the input and output layers, FIGRET uses five fully connected neural network layers with 128 neurons each and employs ReLU(x) activation, except for the output layer, which uses Sigmoid(x).\\
\textbf{Optimizer.} During the training process, FIGRET employs the Adam optimizer \cite{kingma2014adam} for stochastic gradient descent.
\section{Additional link failure results}
\label{section: additional failure results}
\begin{figure}[H]
    \centering
    \includegraphics[scale=0.5]{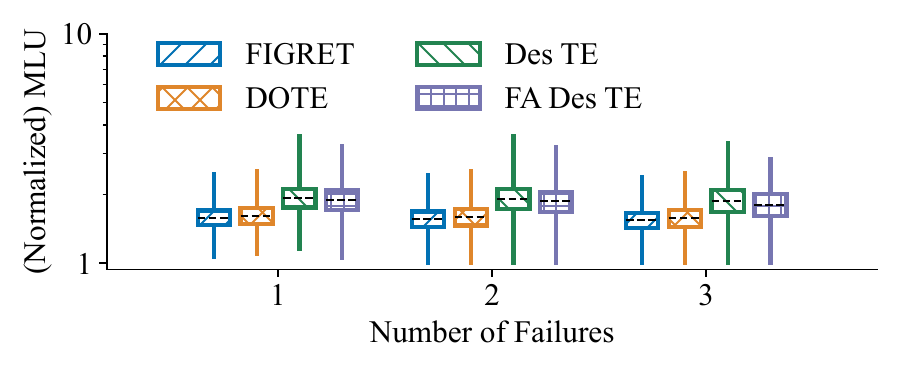}
    \caption{Coping with different numbers of random link failures on pFabric.}
    \label{fig: failure fabric}
\end{figure}

\begin{figure}[H]
    \centering
    \includegraphics[scale=0.5]{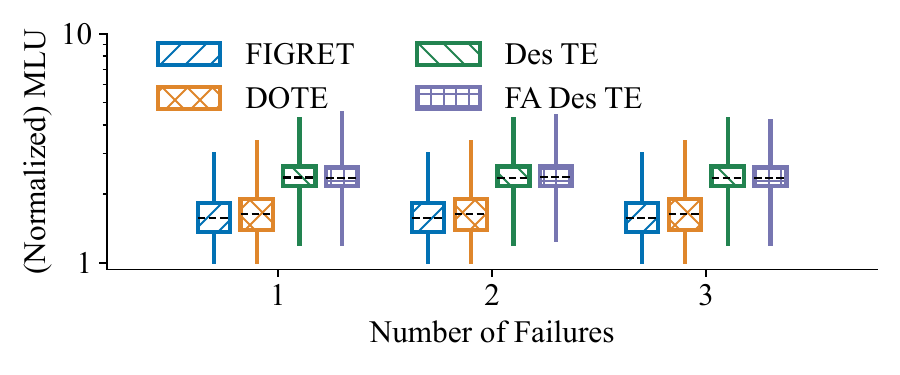}
    \caption{Coping with different numbers of random link failures on ToR-level Meta DB.}
    \label{fig: failure facebook tor a}
\end{figure}

In this section, we present the results from evaluating link failures on pFabric and ToR-level DC Meta DB, as shown in Figures \ref{fig: failure fabric} and \ref{fig: failure facebook tor a}. As can be seen in Figure \ref{fig: failure facebook tor a}, when the network traffic demands exhibit high dynamics, the current TE scheme in Google's Jupiter data centers, even with the knowledge of which links are going to fail, fails to achieve satisfactory results.
\section{Visualization of traffic demands}
\label{section:visualization of traffic demands}
\begin{figure}[!ht]
    \subfigure[0\%-25\%]{
    \label{fig:pod_a_0}
    \includegraphics[scale=0.27]{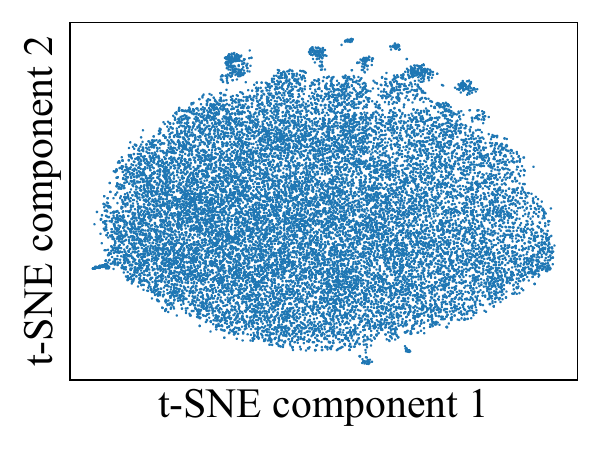}
    }
    \subfigure[25\%-50\%]{
    \label{fig:pod_a_1}
    \includegraphics[scale=0.27]{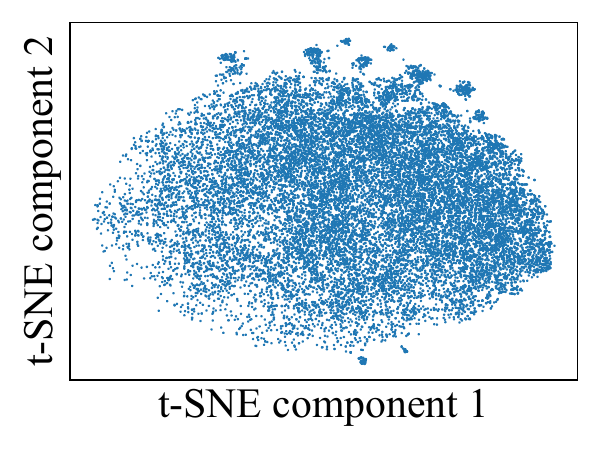}
    }
    \subfigure[50\%-75\%]{
    \label{fig:pod_a_2}
    \includegraphics[scale=0.27]{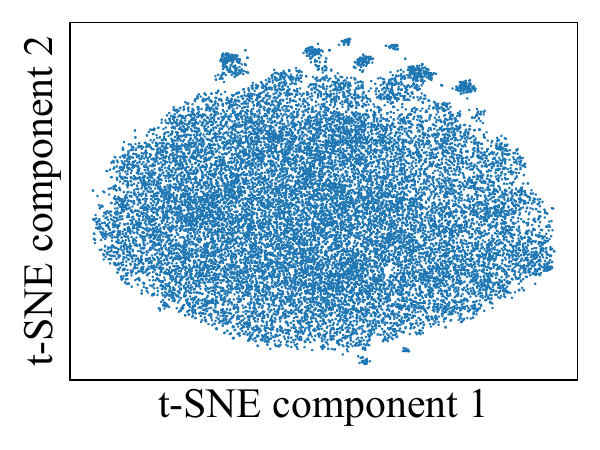}
    }
    \subfigure[75\%-100\%]{
    \label{fig:pod_a_3}
    \includegraphics[scale=0.27]{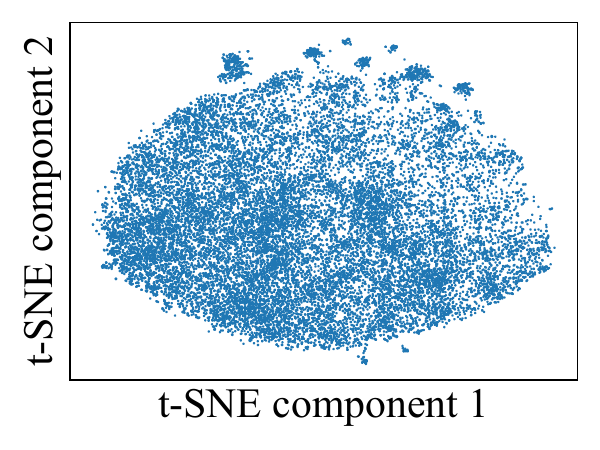}
    }
    \caption{Visualizing Traffic Demands of the PoD-level Meta DB using the t-SNE Method.}
    \label{fig: t-sne facebook pod a}
\end{figure}

\begin{figure}[!ht]
    \subfigure[0\%-25\%]{
    \label{fig:tor_a_0}
    \includegraphics[scale=0.27]{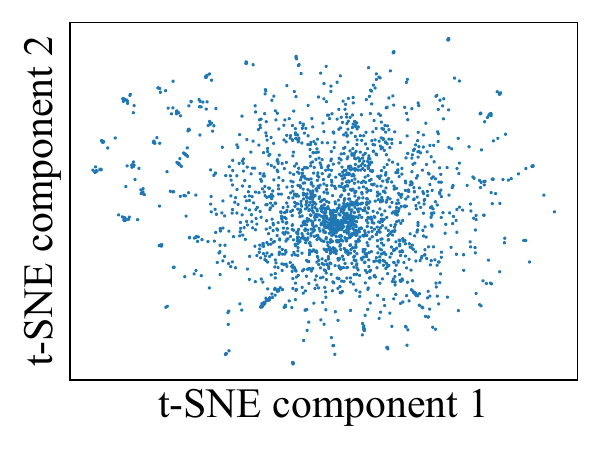}
    }
    \subfigure[25\%-50\%]{
    \label{fig:tor_a_1}
    \includegraphics[scale=0.27]{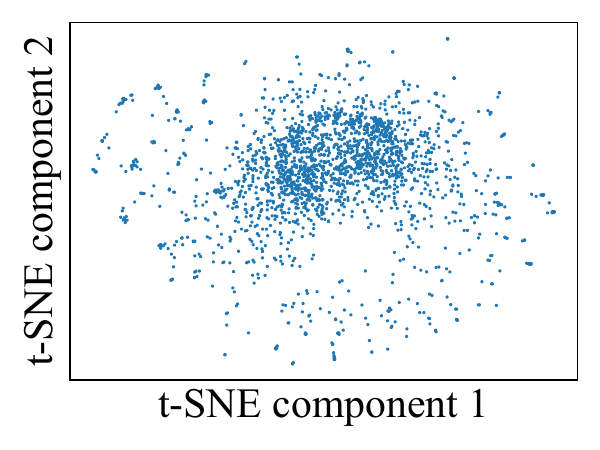}
    }
    \subfigure[50\%-75\%]{
    \label{fig:tor_a_2}
    \includegraphics[scale=0.27]{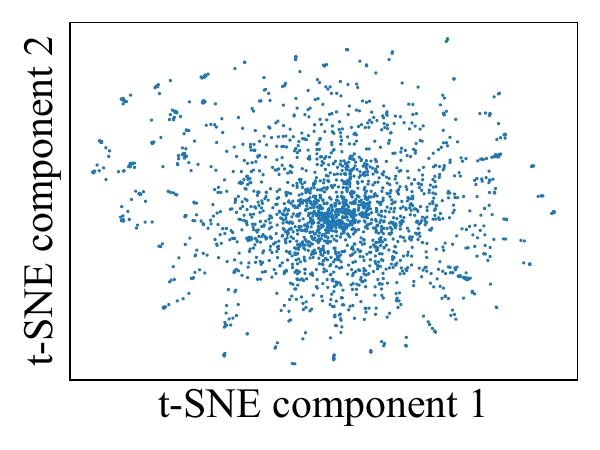}
    }
    \subfigure[75\%-100\%]{
    \label{fig:tor_a_3}
    \includegraphics[scale=0.27]{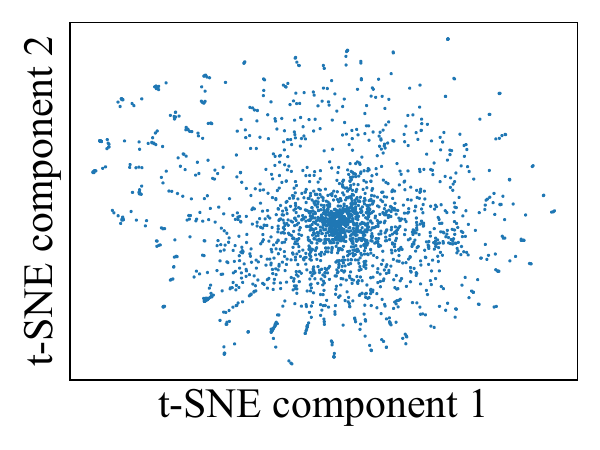}
    }
    \caption{Visualizing Traffic Demands of the ToR-level Meta DB using the t-SNE Method.}
    \label{fig: t-sne facebook tor a}
\end{figure}

To visually demonstrate the extent of traffic data migration over time, we conduct a visualization analysis of traffic demands from both the PoD-level Meta DB and the ToR-level Meta DB using the t-distributed stochastic neighbor embedding (t-SNE) method \cite{hinton2002stochastic} in this section. We analyze their traffic from $0\%$ to $100\%$ using t-SNE and compare the differences in the intervals of $0\%-25\%$, $25\%-50\%$, $50\%-75\%$, and $75\%-100\%$. The 2-dimensional t-SNE components are plotted in Figure \ref{fig: t-sne facebook pod a} and Figure \ref{fig: t-sne facebook tor a}.

From the results, we observe the following:
\begin{itemize}
\item Compared to the PoD-level data, the ToR-level data is more dispersed, indicating a higher dynamism in ToR-level traffic data.
\item Both ToR-level and PoD-level data exhibit a single cluster formation in the t-SNE plots, suggesting their traffic patterns do not undergo drastic changes over time.
\item While the PoD-level data remains very similar across all four-time segments, the ToR-level data shows some variations.
\end{itemize}
These observations are consistent with our evaluation results, indicating firstly that TE is more challenging at the ToR-level (\S \ref{section: comparing figret with other TE schemes}) and, secondly, that FIGRET's robustness to time drift performs well at both levels (\S \ref{section:robustness to demand changes}). Moreover, the impact of time drift on FIGRET's effectiveness is greater at the ToR-level (\S \ref{section:robustness to demand changes}).
\begin{figure*}[ht]
    \subfigure[H = 12]{
    \label{fig:H=12}
    \includegraphics[scale=0.33]{figures/cosine_similarity.pdf}
    }
    \subfigure[H = 64]{
    \label{fig:H=64}
    \includegraphics[scale=0.33]{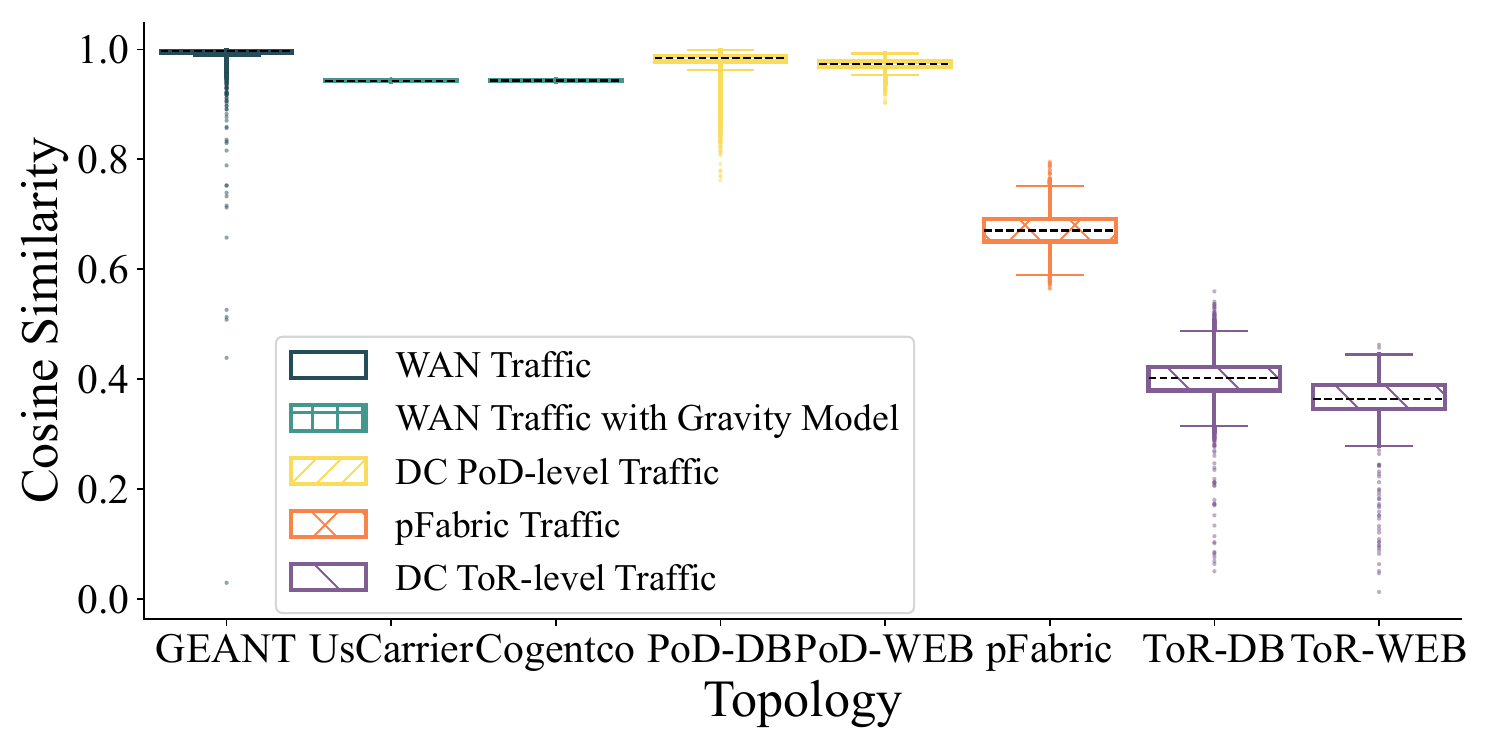}
    }
    \caption{Cosine similarity analysis using a window of H historical TMs vs. the currently-seen TM.}
    \label{fig: window of H}
\end{figure*}
\section{A closer look at the drawbacks of existing TE schemes}
\subsection{Objective mismatch in Demand-prediction-based TE}
\label{section:objective mismatch}
\begin{figure}[!ht]
    \subfigure[In the network topology, where numbers on edges indicate link capacity, traffic demands of \(d_1\) and \(d_2\) need to be sent from \(s\) to \(t_1\) and \(s\) to \(t_2\), respectively.]{
    \label{fig: predict topo}
    \includegraphics[scale=0.4]{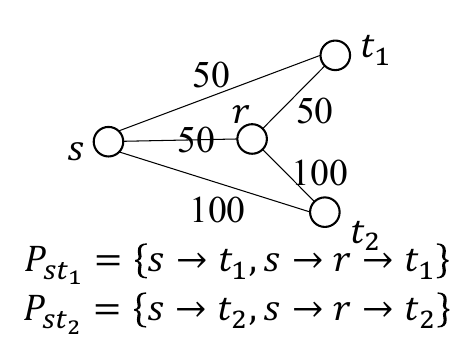}
    }
    \hspace{3mm}
    \subfigure[An example concerning predicted traffic, where the split ratios are only specified for the paths \(s \rightarrow t_1\) and \(s \rightarrow t_2\), with the split ratios for the other two paths being \(1\) minus the corresponding split ratios.]{
    \label{fig: predict table}
    \includegraphics[scale=0.43]{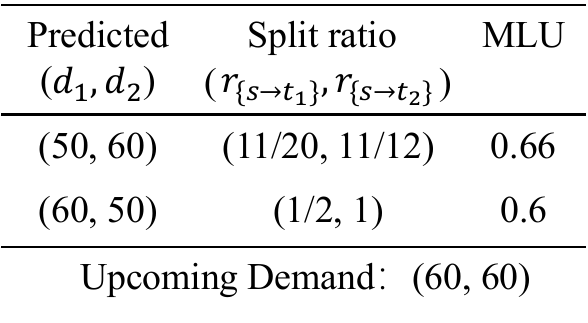}
    }
    \caption{An example demonstrating the mismatch between achieving accurate traffic predictions and minimizing MLU. Two predicted demands achieve the same level of prediction accuracy (with identical mean-square-error values). However, the optimal split ratios derived from these two traffic predictions lead to different MLUs. Incorrect predictions of the traffic from \(s\) to \(t_2\) have a lesser impact on MLU. This is due to the larger capacity of the path from \(s\) to \(t_2\).}
    \label{fig: mismatch in prediction}
\end{figure}

In demand-prediction-based TE, a notable mismatch exists between the objective of accurately predicting future traffic and minimizing the MLU \cite{perry2023dote}. This mismatch primarily arises from the influence of network topology on MLU, indicating that traffic demands exert varied impacts on MLU. For instance, traffic traversing paths with higher capacity tends to have a lesser effect on MLU. Similarly, traffic flowing through paths without shared edges with other paths will also impact MLU to a lesser degree. An example illustrating this concept is provided in Figure \ref{fig: mismatch in prediction}.
\subsection{Limited gains from window expansion}
Deep learning approaches demonstrate powerful capabilities in using information within a window to obtain a better \(D^{\text{expect}}\). Therefore, in this section, we consider whether, under the premise of sufficient computational power and memory, expanding the window size could make \(D^{\text{expect}}\) highly accurate and prevent mis-predictions, thereby eliminating the need to consider sudden traffic changes. Using an analysis method similar to \S \ref{section: evaluation}, we examine the cosine similarity between upcoming traffic and traffic from the past \(H\) time windows, where \(H\) is increased to 64. The results are summarized in Figure \ref{fig: window of H}. 
However, the results indicate that, regardless of whether in stable or bursty network types, the cosine similarity does not significantly increase compared to when \(H\) is 12, suggesting that expanding the window size is \textit{ineffective} in avoiding sudden traffic changes.

\begin{figure}[!h]
    \subfigure[The arrival scenario of the traffic for a SD pair.]{
    \label{fig: dote arrival scenario}
    \includegraphics[scale=0.3]{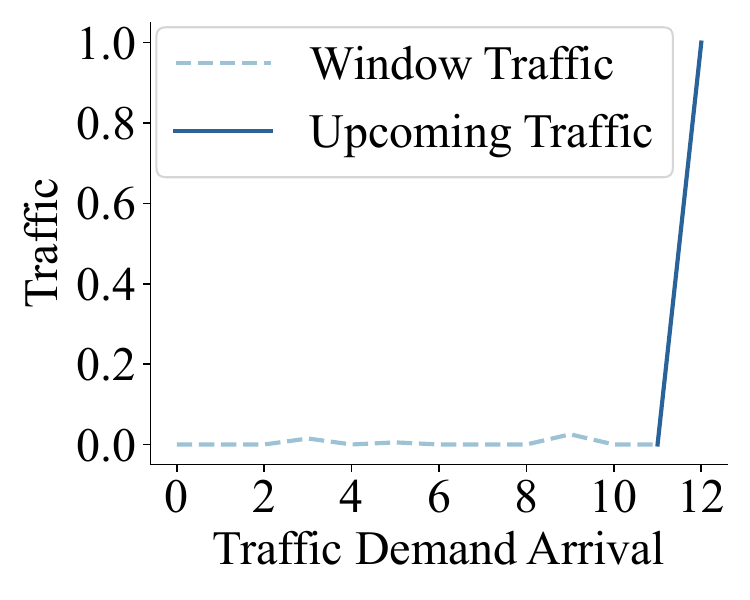}
    }
    \subfigure[Path sensitivity of this SD pair.]{
    \label{fig: dote path sensitivity}
    \includegraphics[scale=0.3]{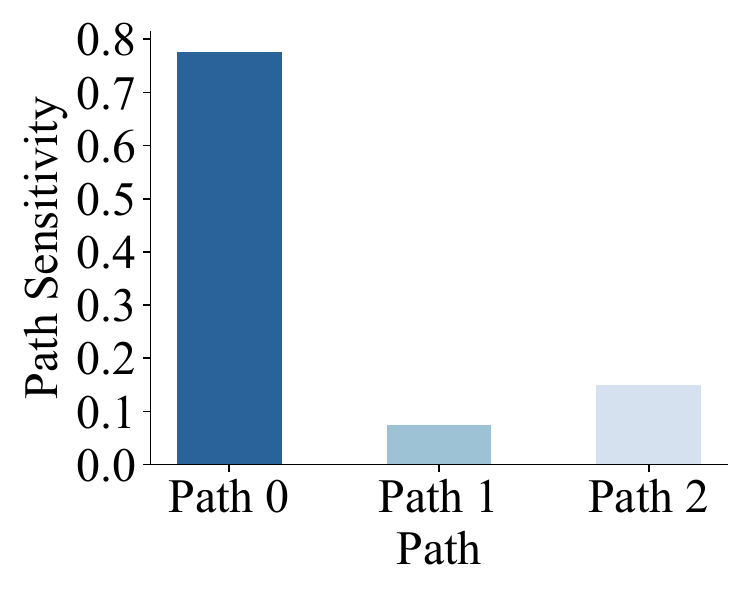}
    }
    \caption{Showcasing the limitations of the DOTE approach.}
    \label{fig: showcasing the limitations of DOTE}
\end{figure}

Due to the limited information within the window, DOTE, which solely relies on window-based information and prioritizes MLU, may result in suboptimal configurations. An example is shown in Figure \ref{fig: showcasing the limitations of DOTE}, which illustrates an inadequate network configuration by DOTE on the ToR-level Meta DB (with an Omniscient-normalized MLU of 2.86). We selected the SD pair responsible for the MLU, and as seen in Figure \ref{fig: dote arrival scenario}, the traffic on this SD pair was consistently stable at a relatively low value. Therefore, DOTE assigned a configuration for this SD pair involving paths with high path sensitivity. However, in the next time snapshot, the SD pair burst unexpectedly, significantly impacting the highly sensitive paths, leading to high MLU.
\end{sloppypar}
\end{document}